\documentclass[12pt,preprint]{aastex}
\shorttitle{Role of rotation and PC--current on pulsar radio emission}  
\shortauthors{D. Kumar and R. T. Gangadhara}
\received{}            
\begin{document}
\title{ROLE OF ROTATION AND POLAR CAP CURRENT ON PULSAR RADIO EMISSION
  AND POLARIZATION} \author{D. Kumar$^1$ and R. T. Gangadhara$^2$}
\affil{Indian Institute of Astrophysics, Bangalore - 560034, India\\
  $^1$dinesh@iiap.res.in, $^2$ganga@iiap.res.in} 
\begin{abstract}
The perturbations such as rotation and PC--current have been believed
to be greatly affecting the pulsar radio emission and polarization.
The two effects have not been considered simultaneously in the
literature, however, each one of these has been considered separately
and deduced the picture by simply superposing them, but such an
approach can lead to spurious results.  Hence by considering pulsar
rotation and PC--current perturbations together instead of one at a
time we have developed a single particle curvature radiation model,
which is expected to be much more realistic.  By simulating a set of
typical pulse profiles we have made an attempt to explain most of the
observational results on pulsar radio emission and polarization.  The
model predicts that due to the perturbations leading side component
can become either stronger or weaker than the corresponding trailing
one in any given cone depending on the passage of sight line and
modulation (nonuniform source distribution). Further we find that the
phase delay of polarization angle inflection point with respect to the
core component greatly depends upon the viewing geometry. The
correlation between the sign reversal of circular polarization and the
polarization angle swing in the case of core dominated pulsars become
obscure once the perturbations and modulation become significant.
However the correlation that the negative circular polarization
associates with the increasing polarization angle and vice versa shows
up very clearly in the case of conal$-$double pulsars.  The `kinky'
type distortions in polarization angle swing could be due to the
incoherent superposition of modulated emissions in the presence of
strong perturbations.
\end{abstract} 

\section{INTRODUCTION}
Pulsars which are famed by their highly periodic signals are now
universally accepted as fast rotating and highly magnetized (mainly
dipolar) neutron stars. The coherent curvature radiation due to the
ultra-relativistic plasma streaming out along the open dipolar
magnetic field lines is believed to be responsible for the pulsar
radio emission \citep[e.g.,][]{S1971,RS1975,MGP2000,GLM2004}.  The
individual pulses from pulsars in general are highly random in
strength as well as their appearance in longitude within the pulse
window. However the average profiles resulted from the summation of
several hundreds of individual pulses have well defined shapes, and
they are unique in most of the cases.  Further pulsars in general show
a ``S'' shaped characteristic polarization position angle (PPA) swing
which is attributed to the underlying dipole field geometry of
emission region \citep{RC1969}.

The average profiles in general are made up of many components.
\citet{R1983,R1990,R1993}, \citet{MD1999} and \citet{MR2002} have
recognized that the pulsar emission beams have nested core-cone
structure. But the conal components in general show asymmetry in their
location with respect to the central core component. Hence
\citet{LM1988} have suggested that the emission is `patchy'.  The
components often show  asymmetry in their strengths between the
leading and trailing sides of the profiles. Further, some pulsars show
the polarization angle that deviates from the standard `S'
curve \citep{Xetal1998}.

Among the several relativistic effects that have been proposed to
understand pulsar emission and polarization, the effects of rotation
such as aberration and retardation (A/R) and polar-cap current
(PC-current) perturbation are found to be quite important. Due to
pulsar rotation the relativistic plasma gets corotation velocity
component which is in addition to the intrinsic velocity along the
dipole field lines.  Hence the net velocity of plasma will be
aberrated in the direction of pulsar rotation. Therefore, an inertial
observer tend to see the plasma trajectory which differs significantly
from the associated dipole field lines, and hence affecting the pulsar
emission and polarization (Blaskiewicz, Cordes and Wasserman 1991,
hereafter BCW 1991; Dyks 2008; Thomas \& Gangadhara 2007; Thomas \&
Gangadhara 2010; Dyks et~al. 2010; Thomas et al. 2010; Kumar \&
Gangadhara 2012a, hereafter KG 2012a; Wang et al. 2012). On the other
hand, in yet another artificial models where the corotation of the
pulsar magnetosphere is ignored and considered only the effect of
PC--current on the underlying dipole field. The field lines will get
curvature in the azimuthal direction due to the PC--current induced
toroidal magnetic field which is in addition to their intrinsic
curvature in the polar direction. Therefore, the trajectory of field
line constrained plasma becomes significantly different from the
unperturbed case and hence affect the pulsar emission and polarization
(Hibschman \& Arons 2001; Gangadhara 2005; Kumar \& Gangadhara 2012b,
hereafter KG 2012b).

By taking into account of rotation, \citet{BCW1991} have predicted that the PPA inflection point lags the
midpoint of the intensity profile by $\sim 4 r/r_{LC},$ where
$r_{LC}=cP/2 \pi$ is the light cylinder radius, and $c$ is the
velocity of light and $P$ is the pulsar rotation period. Later \citet{D2008} have confirmed this behavior. However,
\citet{KG2012a} have shown that due to the combined effect of rotation,
modulation and viewing geometry the phase lag of the PPA inflection
point with respect to the central core will become significantly different
from $\sim 4 r/r_{LC}.$ On the other hand, \citet{KG2012b}
have predicted that the PPA inflection point can even lead the central
core due to the PC--current-induced perturbation. Note that an
asymmetry in the phase location of components is believed to arise from
the aberration and retardation effects \citep[]{GG2001,GG2003,DRH2004,G2005}.

Further, by considering the rotation \citet{BCW1991} have predicted that the leading side intensity
components dominate over the trailing ones. This is due to the fact
that the curvature of source trajectory on leading side becomes larger
than that on the trailing side. Later \citet{TG2007} have confirmed this effect and \citet{DWD2010}, \citet{Tetal2010}, \citet{KG2012a} and
\citet{Wetal2012} have reconfirmed this behavior. Although, statistically the cases with
leading component stronger are more common, the converse cases are
also quite significant \citep{LM1988}. By considering the
PC--current-induced perturbation on the underlying dipole field
\citet{KG2012b} have shown that the leading side components can get
either stronger or weaker than the corresponding ones on trailing side
depending upon the viewing geometry and modulation. However the
aforementioned prediction by \citet{KG2012b} was in the absence of
strong rotation effect.

In literature two types of circular polarization have been recognized:
`antisymmetric' and `symmetric'. If the circular polarization changes
its sense near the center of the pulse profile then it recognized as
the antisymmetric$-$type circular polarization whereas if the polarity
of the circular polarization does not change through out the pulse
profile then it recognized as the symmetric$-$type
\citep{RR1990}. However, either antisymmetric or symmetric circular
polarization can be associated with the individual components
\citep{Hanetal1998,YH2006,KG2012a,KG2012b}. Earlier only the
antisymmetric--type circular polarization was thought to be an
intrinsic property of curvature radiation
\citep[e.g.,][]{M1987,GS1990a,GS1990b,RR1990,Getal1993,G1997,G2010}
and the origin of `symmetric'$-$type circular polarization was
speculated to be through propagation effects. Recently \citet{KG2012a}
have shown however that in addition to the antisymmetric, the
symmetric circular polarization can also be produced within the
framework of curvature radiation if the effects of pulsar rotation,
nonuniform plasma distribution and viewing geometry are taken into
account. Recently \citet{Wetal2012} have also reconfirmed the findings
of \citet{KG2012a}. \citet{KG2012b} deduced the similar result by
considering PC-current induced perturbation instead of pulsar
rotation.

Pulsars show a diverse behavior in circular polarization among which
its association with the PPA swing is quite important for
understanding the underlying geometry of emission region. In the case
of pulsars with antisymmetric circular polarization, \citet{RR1990}
have found a strong correlation between the sense reversal of circular
polarization and the PPA swing: the sign reversal of circular
polarization from negative to positive is associated with the
increasing PPA sweep and vice versa. \citet{G2010} has confirmed the
correlation and proposed it as a geometric phenomenon. Further
\citet{KG2012a,KG2012b}, from their simulation of polarization
profiles, showed that the correlation exists when the rotation and
PC-current perturbations are less significant. But
\citet{Hanetal1998},\citet{YH2006} have noticed that the sense
reversal of circular polarization near the center of pulse profiles is
not correlated with the PPA swing. However, they did find a strong
correlation between the sense of the circular polarization and the PPA
swing in double$-$conal pulsars. \citet{KG2012a} speculated that such
a correlation in the case of double$-$conal pulsars can arise if the
modulations are asymmetrically located in the conal rings centered on
the magnetic axis.

Among the several relativistic models proposed by taking into account
of pulsar rotation and PC-current perturbations, only the models
proposed by \citet{KG2012a}, \citet{Wetal2012} and \citet{KG2012b} can
explain full polarization state of the radiation field. Further they
are much more realistic in the sense that in addition to strong
perturbations the effects of nonuniform source distribution
(modulation) and viewing geometry have been incorporated as an
essential ingredients. However in the models proposed by
\citet{KG2012a} and \citet{Wetal2012} as a special case considered the
effect of pulsar rotation and ignored PC--current perturbation. On the
other hand, in yet another artificial case \citet{KG2012b} considered
the effect of PC--current on the underlying dipole field by ignoring
the corotation of the pulsar magnetosphere. Since both the effects are
found to be quite dominant in affecting pulsar radio emission and
polarization, they have to be combined together.

In this paper both the rotation and PC--current perturbations are
considered simultaneously, and analyzed their combined effect on the
pulsar emission. If we consider separately the rotation and
PC--current perturbations, and add up the results, the resulting
conclusions can become erroneous as we show in next section and hence
this work is an important one. Although pulsar radiation is generated via some coherent processes, we perform the modeling of pulsar radio emission in terms of single particle curvature radiation. Note that although to the first order the single particle approximation is not a bad assumption, but in reality some factors influencing the coherence processes may favor or oppose the effects that we consider in this work.  We present the theory of single particle curvature
radiation in a rotating PC--current perturbed magnetic field and
analyze the polarization state in section~2.  In section~3, we present
a set of simulated pulse profiles, and speculate on the polarization
properties by comparing the observed pulses.  In section~4, we give
the discussions and in section~5 make the conclusion. 

\section{CURVATURE RADIATION FROM ROTATING PC--CURRENT PERTURBED MAGNETOSPHERE}
\begin{figure}
\centering
\epsscale{0.9}
\plotone{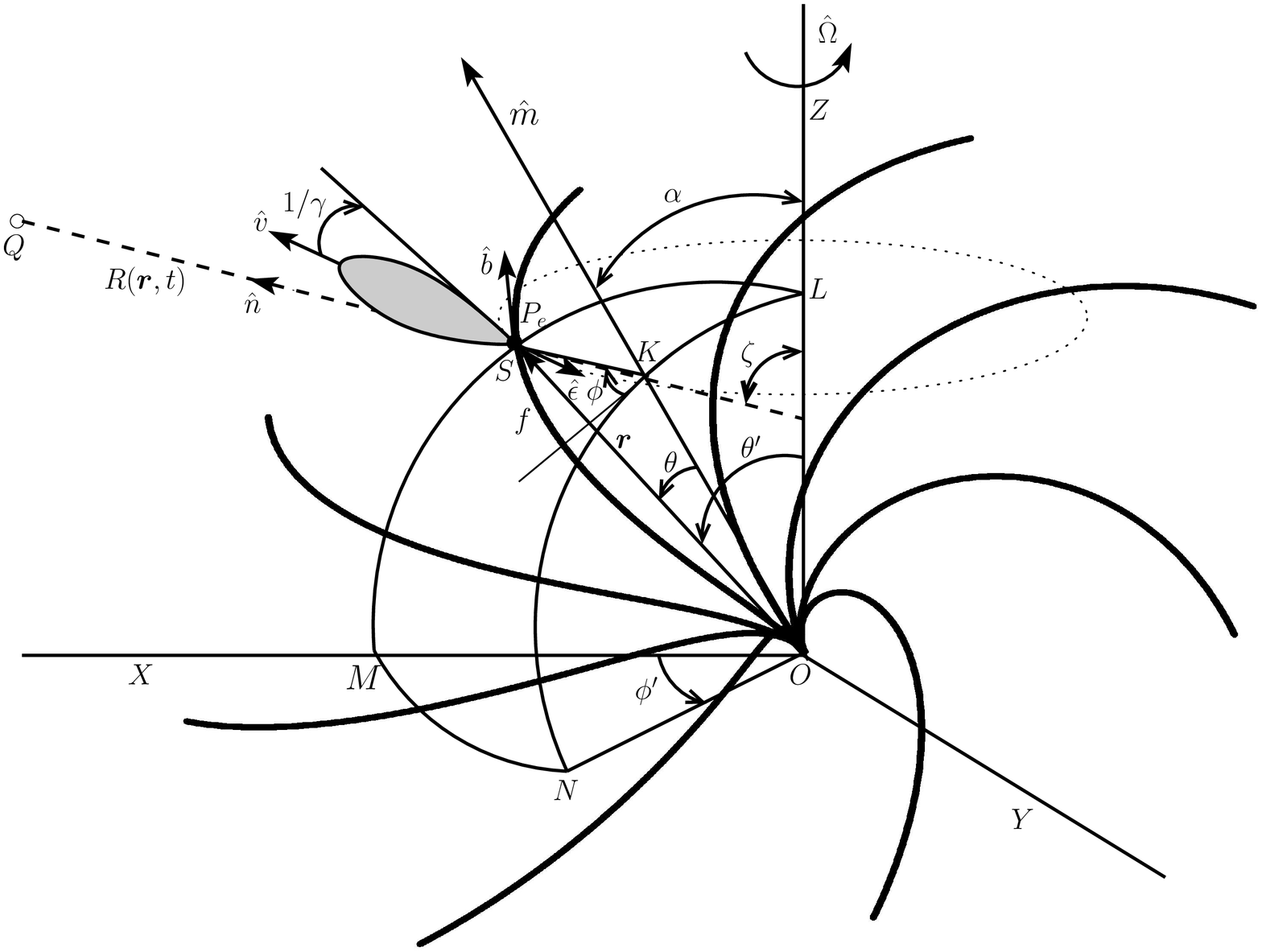}
\caption{Geometry of curvature radiation in a rotating
  PC--current-perturbed magnetic dipole with the Cartesian coordinate
  system$-XYZ$ as an inertial observer's frame whose origin is the
  neutron star center at $O.$ The magnetic axis $\hat m$ is inclined
  to the rotation axis $\hat \Omega$ by an angle $\alpha,$ and the
  rotation phase $\phi'$ of the magnetic axis is measured from the
  fiducial $XZ-$ plane. The thick solid line curves represent the
  PC--current-perturbed field lines, and the parameters
  $\alpha=30^\circ,$ $\phi'=30^\circ,$ field line constant $r_e=1
  r_{LC},$ pulsar rotation period $P=1$ s, a current scale factor
  $\varsigma=1,$ magnetic azimuth $\phi_{i}$ from $0^\circ$ to
  $360^\circ$ in steps of $45^\circ$, magnetic colatitude
  $\theta_{i}=0^\circ$, and emission altitude $r$ from $0$ to
  $0.8~r_{\rm LC}$ are used to sketch them. The unit vectors $\hat b$
  and $\hat \epsilon$ represent the direction of perturbed field line
  tangent and pulsar rotation, respectively. The source net velocity
  $\hat v$ represent the direction of beamed radiation.}
 \label{fig:Figure1}
\end{figure}

Let us consider a stationary Cartesian coordinate system--XYZ with the
origin~O located at neutron star center as an inertial observer frame
(see Figure~\ref{fig:Figure1}). Consider an inclined and rotating
PC--current perturbed magnetic dipole with an inclination angle
$\alpha$ with respect to the rotation axis $\hat \Omega,$ which is
taken to be parallel to Z-axis. The velocity $\mbox{\boldmath $v$}$ of
the relativistic source $S,$ which is constrained to move along the
rotating PC--current perturbed dipole field line f, is given by
\begin{eqnarray}
\mbox{\boldmath $v$} =  \kappa c \,\hat{b} + \mathbf\Omega\times\textbf{r}~,
\label{eqn:v}
\end{eqnarray}
where $\hat{b}=\mbox{\boldmath $B$}/|\mbox{\boldmath $B$}|$ and
$\mbox{\boldmath $B$}=\mbox{\boldmath $B_0$}+\mbox{\boldmath $B_1$}.$
And $\mbox{\boldmath $B_0$}$ is the unperturbed dipole field,
$\mbox{\boldmath $B_1$}$ is the PC--current-induced field, and
$\textbf{r}$ is the position vector of the source \citep{KG2012b}. The
parameters $\mathbf\Omega=\Omega \hat\Omega$ is the pulsar angular
velocity and $\kappa$ specifies the normalized speed of the source
with respect to the speed of light c along the associated perturbed
field line.

The first term on the r.h.s.~of Equation~(\ref{eqn:v}) is the velocity
of source along the perturbed field lines. The second term is the
induced velocity due to corotation of the pulsar magnetosphere. Note
that due to the PC--current-perturbation, the field lines which lie
above the magnetic axis, tend to azimuthally twist towards the pulsar
rotation, whereas those which lie below the magnetic axis, twist in
the opposite directions \citep[see Figure 1 in][]{KG2012b}.
Therefore, the contributions to aberration of the source velocity
$\mbox{\boldmath $v$}$ by the above two terms add up for the negative
sight line impact angle $\sigma,$ but they try to cancel each other
for the positive $\sigma.$ However, since the aberration of source
velocity due to the effect of rotation is larger than that due the
PC--current-perturbation for the current which is of order
Goldreich-Julian current, the net aberration will be always in the
direction of pulsar rotation, and it is greater for negative $\sigma.$

Since the relativistic emissions are beamed in the direction of
velocity $\mbox{\boldmath $v$}$ with half-opening angle $1/\gamma,$
observer can receive the emissions only from a selected emission
region whose boundary satisfies $\hat n \cdot \hat v=\cos (1/\gamma),$
where $\hat v=\mbox{\boldmath $v$}/|\mbox{\boldmath $v$}|$ and the
sight line $\hat n=\left\lbrace \sin\zeta,~0,~\cos\zeta\right\rbrace$
with $\zeta=\alpha + \sigma.$ But the exact analytical solutions for
the emission point coordinates $(\theta_{0},~\phi_{0})$ and
$(\theta_{e},~\phi_{e})$ (see \citet{KG2012a} and \citet{KG2012b} for
their definition) of the beaming region are hard to find once the
effects of rotation and PC--current perturbation are considered, and
hence we seek numerical solutions.  Note that in finding exact values
for coordinates $\theta_{0}$ and $\phi_{0},$ the values obtained from
\citet{KG2012b} have to be used as initial guess values for fast
convergence unlike \citet{KG2012a} wherein they have used those
derived from \cite{G2010}.

\begin{figure}
\centering
\epsscale{1.0}
\plotone{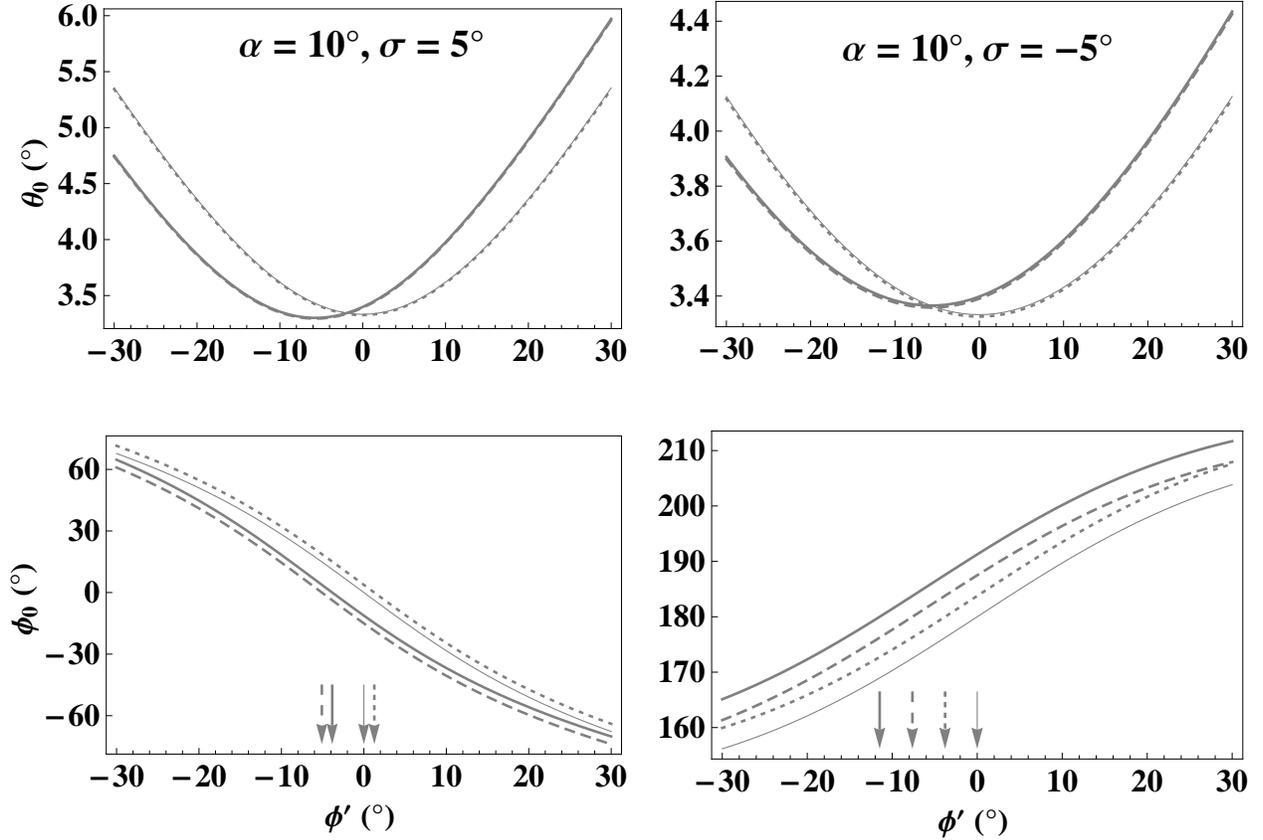}
\caption{Emission point coordinates $\theta_0$ and $\phi_0$ as
  functions of rotation phase $\phi'$ in the cases of nonrotating
  dipole (thin solid line curves), nonrotating PC--current-perturbed
  dipole (dotted line curves), rotating dipole (dashed line curves),
  and rotating PC--current-perturbed dipole (thick solid line
  curves). The arrows in the $\phi_0$ panels (line styled similar to
  $\phi_0$) indicate the antisymmetric point of $\phi_0$ in the
  respective cases. Here we used $r_n=0.1,$ $P=1$ s, $\gamma=400,$ and
  $\varsigma=1.$}
 \label{fig:Figure2}
\end{figure}

By using the parameters $r_n=0.1,$ $P=1$ s, source Lorentz factor
$\gamma=400,$ the current scale factor $\varsigma=1$,
$\alpha=10^\circ,$ and $\sigma=\pm5^\circ,$ we computed $\theta_{0}$
and $\phi_{0},$ and presented in Figure~\ref{fig:Figure2} as function of
$\phi'.$ Due to the PC--current perturbation alone (represented by
dotted line curves), the emission points in $\phi_{0}$ shift to later
phases for positive $\sigma$ and to the earlier phases for negative
$\sigma,$ whereas they are mostly unaffected in $\theta.$ On the other
hand due to the effect of rotation (aberration) alone (represented by
dashed line curves), the emission shifts to earlier phases in both
$\theta$ and $\phi.$ However the phase shift of emission points in
$\phi$ caused due to rotation alone is larger than that due to
PC--current perturbation alone. This can be clearly seen in the phase
shifts of the antisymmetric point of $\phi_0$ (phase at which $\phi_0$
is equal to $0^\circ$ or $180^\circ$) indicated by arrows (styled same
as $\phi_0$). As a result, in the more realistic case of rotating
PC--current perturbed dipole (represented by thick solid line curves),
the emission points in $\theta$ shift to earlier phases in both the
cases of $\sigma$ by the same amount as that in the case of rotating
dipole, whereas in $\phi,$ they shift to the earlier phases by a
smaller amount in the case of positive $\sigma$ and by larger amount
in the case of negative $\sigma.$ Note that, if the emission region is
modulated (nonuniform source distribution) in the azimuthal direction
which we show later, then the resulted intensity components will also
show aforesaid asymmetric phase shift between the positive and
negative $\sigma$ cases.

We also computed $\theta_0$ and $\phi_0$ for the case of rotating
perturbed dipole by assuming $\theta_0=\theta_0'+\delta\theta_{\rm
  rot}+\delta\theta_{\rm pcc},$ and $\phi_0=\phi_0'+\delta\phi_{\rm
  rot}+\delta\phi_{\rm pcc},$ where $\theta_0'$ and $\phi_0'$ are the
coordinates of emission points in the nonrotating unperturbed
dipole. The changes in the colatitude $\delta\theta_{\rm
  rot}=\theta_{0,\rm rot}-\theta_0'$ and $\delta\theta_{\rm
  pcc}=\theta_{0,\rm pcc}-\theta_0'$ and those in the azimuth
$\delta\phi_{\rm rot}=\phi_{0,\rm rot}-\phi_0'$ and $\delta\phi_{\rm
  pcc}=\phi_{0,\rm pcc}-\phi_0'$ are due to the aberration and
PC--current perturbation, respectively. The parameters $\theta_{0,\rm
  rot}$ and $\theta_{0,\rm pcc}$ are the coordinates after separately
considering the perturbations: rotation and PC--current, respectively,
and similarly $\phi_{0,\rm rot}$ and $\phi_{0,\rm pcc}.$ We find that
thus obtained $\theta_0$ and $\phi_0$ more or less match with the
thick solid line curves (not shown in the Figure). Therefore, the
phase shifts as well as the changes in the magnitude of emission point
coordinates $\theta_0$ and $\phi_0$ due to the two separate
perturbations (rotation and PC--current), simply add up when the two
effects are combined together.

The acceleration  of source is given by  
\begin{eqnarray} 
\mbox{\boldmath $a$} =\frac{(\kappa
  c)^{2}}{|\mbox{\boldmath $b$}|}\frac{d\hat{b}}{d\theta} +
\frac{\kappa
  c^{2}}{|\mbox{\boldmath $b$}|}\frac{d\kappa}{d\theta}\hat{b}+2
\kappa c (\mbox{\boldmath $\Omega$}\times\hat{b}) + \mbox{\boldmath
  $\Omega$}\times(\mbox{\boldmath $\Omega$}\times\textbf{r})~,
\label{eqn:a}
\end{eqnarray}
where we have used the expression of arc length of the field line $ds
= |\mbox{\boldmath $b$}| d\theta = \kappa c ~dt,$ and the total
derivative $dF/d\theta=\partial F/\partial\theta+(\partial
F/\partial\phi) (d\phi/d\theta),$ where $F$ stands for $\hat{b},$
$\kappa,$ etc. The first term on r.h.s.~of Equation~(\ref{eqn:a}) is
the acceleration of bunch due to curvature of the PC--current
perturbed field lines. Note that it includes both the intrinsic
curvature due to the dipolar field and the induced curvature due the
PC-current perturbation on the dipole field. The small change in the
source speed due to motion along the perturbed field line is
represented by the second term and it is tiny among other terms. The
third term is the rotationally induced acceleration due to the
Coriolis force and is in the direction of pulsar rotation. The last
term is the induced acceleration due to the Centrifugal force which is
acting away from the rotation axis. The PC--current-induced
acceleration which is included in the first term becomes much
important at higher emission altitude, larger pulsar angular velocity,
and smaller $\alpha.$ On the other hand, the Coriolis and the
centrifugal accelerations with the Coriolis term being the dominant
one become much important at higher emission altitude, larger pulsar
angular velocity, but at larger $\alpha.$

\begin{figure}
\centering
\epsscale{1.0}
\plotone{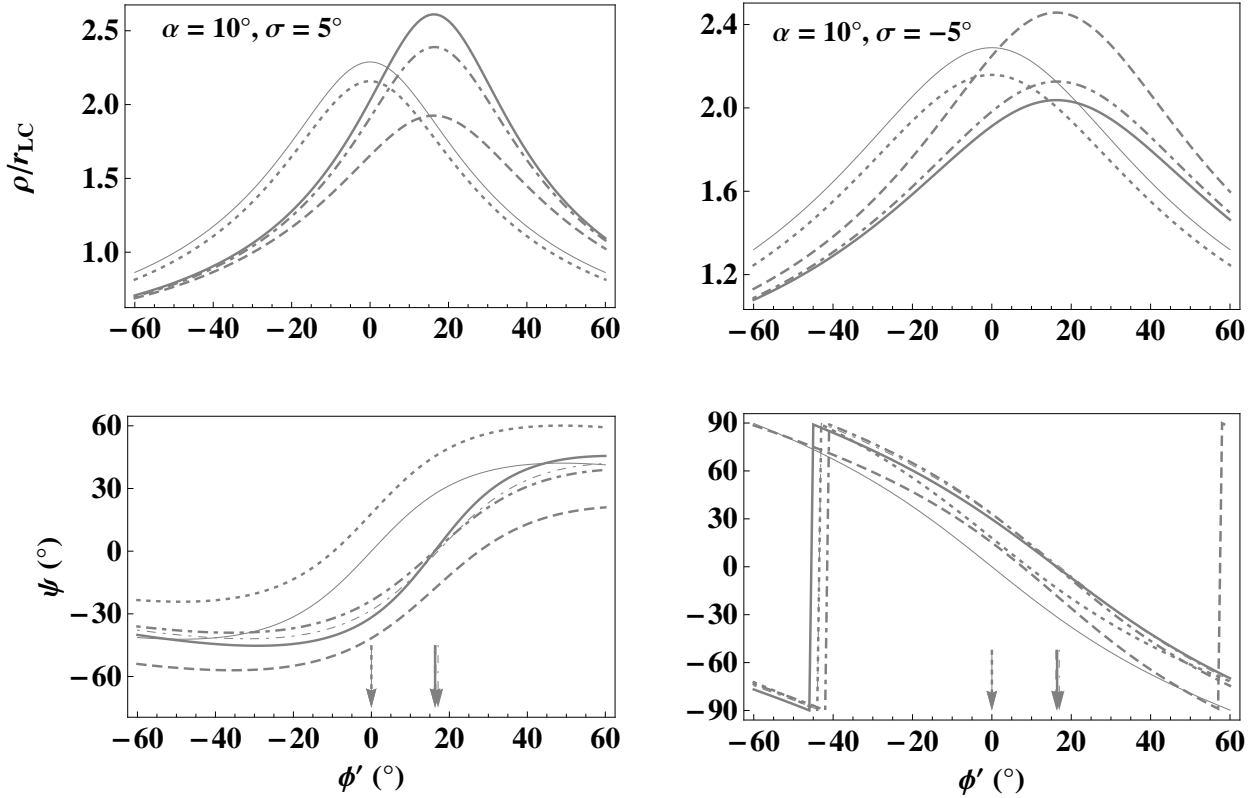}
\caption{Radius of curvature $\rho$ of source trajectory, and position
  angle $\psi$ of the radiation electric field as functions of
  $\phi'.$ The lines representation is the same as in Figure
  \ref{fig:Figure2}. The dot-dashed line curves represent the
  parameters obtained by simple addition of the two perturbations due
  to the rotation and the PC--current. The thin dot-dashed line curve in
  the $\psi$ panels represents the analytical perturbation theory
  \citep{HA2001} whereas the thick dot-dashed line curve represents our
  numerical result. The arrows in the $\psi$ panels (styled same as
  $\psi$) represent the inflection point of $\psi.$ For simulation we
  used the parameters which are the same as in Figure
  \ref{fig:Figure2}.}
 \label{fig:Figure3}
\end{figure}

By using the parameters, which are the same as in Figure
\ref{fig:Figure1}, we computed the radius of curvature $\rho \approx
v^3/|\mbox{\boldmath $v$} \times \mbox{\boldmath $a$}|$ as a function
of $\phi'$, and plotted in upper panels of Figure
\ref{fig:Figure3}. The PC--current perturbation (represented by the
dotted line curves) leads to larger curvature with respect the
unperturbed ones (thin solid line curve) leaving the symmetric point
of $\rho$ to be mostly unaffected \citep{KG2012b}. On the other hand,
rotation induces an asymmetry into the curvature between the leading
and trailing sides of $\phi'=0^\circ,$ since the leading side
trajectory becomes more curved and $\rho$ maximum shifts to the
trailing side \cite[e.g.,][]{KG2012a}. Since the perturbation due to 
PC--current in the case of positive $\sigma$ opposes that due to
the corotation, the net $\rho$ in the rotating PC--current perturbed
dipole (thick solid line curve) becomes larger than that due to
rotation alone, and vice versa in the case of negative $\sigma.$

To assess the possibility of deriving $\rho$ for the rotating
PC--current perturbed dipole by considering separately the
perturbations due to PC--current and rotation, analogues to the
$\theta_0$ and $\phi_0$ in Figure~\ref{fig:Figure2}, we considered the
net curvature as $\mbox{\boldmath$k$}
=\mbox{\boldmath$k'$}+\mbox{\boldmath$\delta k_{\rm
    rot}$}+\mbox{\boldmath$\delta k_{\rm pcc}$},$ where
$\mbox{\boldmath$k'$}$ is the curvature of the unperturbed dipolar
field lines, $\mbox{\boldmath$\delta k_{\rm
    rot}$}=\mbox{\boldmath$k_{\rm rot}$}-\mbox{\boldmath$k'$}$ and
$\mbox{\boldmath$\delta k_{\rm pcc}$}=\mbox{\boldmath$k_{\rm
    pcc}$}-\mbox{\boldmath$k'$}.$ The $\mbox{\boldmath$k_{\rm rot}$}$
and $\mbox{\boldmath$k_{\rm pcc }$}$ are the curvature vectors in the
cases of rotating dipole and PC--current perturbed dipole,
respectively. The resultant $\rho$ can be derived as
\begin{equation}
\rho=\left[\frac{1}{\rho'^2}+\frac{1}{\rho_{\rm
      rot}^2}+\frac{1}{\rho_{\rm pcc}^2}-2\left(\frac{\hat a'\cdot
    \hat a_{\rm rot}}{\rho'~\rho_{\rm rot}}+\frac{\hat a'\cdot \hat
    a_{\rm pcc}}{\rho'~\rho_{\rm pcc}}-\frac{\hat a_{\rm rot}\cdot
    \hat a_{\rm pcc}}{\rho_{\rm rot}~\rho_{\rm
      pcc}}\right)\right]^{-1/2},
\label{eqn:rho}
\end{equation}
where $\rho',$ $\rho_{\rm rot},$ and $\rho_{\rm pcc}$ are the radii of
curvature and $\hat a',$ $\hat a_{\rm rot},$ and $\hat a_{\rm pcc}$
are the unit acceleration vectors in the cases of non-rotating dipole,
rotating dipole and PC--current perturbed dipole, respectively. Thus
obtained $\rho,$ by adding the two separate perturbations computed
independently, is superposed in $\rho$ panels of Figure
\ref{fig:Figure3} (see the dot-dashed line curves). We can see that, it
is significantly different from the actual $\rho$ of the rotating
PC--current perturbed dipole.

The polarization position angle $\psi$ of the electric field of
radiation due to relativistic sources, defined as the angle between
the radiation electric field and the projected spin axis on the plane
of the sky, can be computed by knowing the acceleration of the
radiation source: $\tan\psi=\hat \epsilon_\perp \cdot \hat a / (\hat
\epsilon_\parallel \cdot \hat a),$ where $\hat \epsilon_\parallel$
(projected spin vector) and $\hat \epsilon_\perp$ are unit vectors in
the directions perpendicular to $\hat n$ \citep{G2010}. By using the
parameters which are same as in upper panels of Figure
\ref{fig:Figure3}, we computed the position angle $\psi$ as a function
of $\phi',$ and plotted in lower panels of Figure
\ref{fig:Figure3}. The perturbation due to the PC--current
(represented by dotted line curves) causes the $\psi$ curve to shift
upward with respect to the standard RVM curve (nonrotating dipole,
thin solid line curve) while its inflection point remain unaffected
\citep{HA2001,KG2012b}. On the other hand the rotation (aberration)
causes the $\psi$ curve to shift upward or downward, depending upon
sign of $\sigma$, and always shifts the inflection point to the
trailing side. Therefore in a more realistic case of rotating
PC--current perturbed dipole the net $\psi$ will be shifted upward
with respect to the one in rotating case, while the inflection point
lies mostly at the same phase as in the rotating dipole.

We have also computed $\psi$ for the rotating perturbed dipole by
assuming $\psi=\psi'+\delta\psi_{\rm rot} +\delta\psi_{\rm pcc},$
where $\psi'$ is due to the nonrotating dipole, $\delta\psi_{\rm
  rot}=\psi_{\rm rot}-\psi'$, and $\delta\psi_{\rm pcc}=\psi_{\rm
  pcc}-\psi'$ with $\psi_{\rm rot}$ and $\psi_{\rm pcc}$ are being the
position angles after considering the perturbation separately due to
rotation and PC--current, respectively. We find that thus obtained
$\psi$ in both numerical (represented by thick dot-dashed line curves)
and analytical perturbation theory \citep[represented by thin
  dot-dashed line curves, see the Equations (D13) and (G11) in][]{HA2001}
significantly differ from the actual $\psi$ of the rotating
PC--current-dipole. Therefore we believe that the two effects have to
be combined together in deducing the polarization state of the
radiation field.

\section{POLARIZATION STATE OF THE RADIATION FIELD}
The radiation emitted by the relativistic accelerating sources will
have a broad spectrum, and it is given by \citep{Jackson1998}:
\begin{equation}
\textbf{E}(\textbf{r},\omega) = \frac{1}{\sqrt{2 \pi}}\frac{qe^{i \omega
    R_{0}/c}}{R_{0}~c}\int^{+\infty}_{-\infty}
\frac{|\textbf{b}|}{\kappa c}\frac{\hat{n}\times[(\hat{n} -
    \mbox{\boldmath $\beta$})\times\mbox{\boldmath
      $\dot{\beta}$}]}{\xi^{2}} e^{i \omega (
  t-\hat{n}\cdot \textbf{r}/c)} d\theta.
\label{eqn:Ew} 
\end{equation} 
Note that the time $t$ in the above equation has to be replaced by the
expression given in Equation (6) of \citet{KG2012a}, and the
parameters $\phi'$ by $\Omega~t$ and $\phi$ by the expression given in
Equation (11) of \citet{KG2012b}. We solve the integral using the method given in \cite{KG2012a}
and find the polarization state of the radiation field in terms of the
Stokes parameters $I,$ $Q,$ $U$ and $V.$
\subsection{Emission from Uniform Distribution of Sources}
\begin{figure}
\centering
\epsscale{0.65}
\plotone{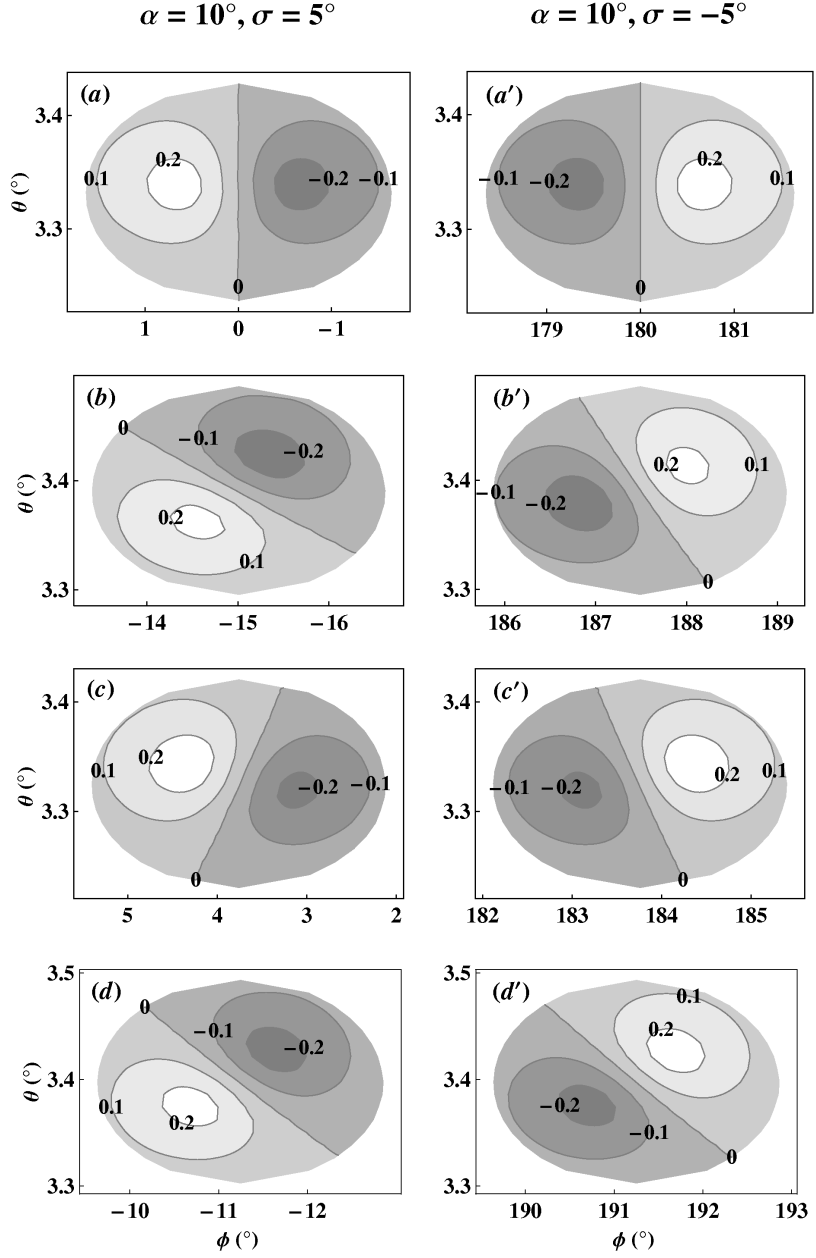}
\caption{The pattern of the circular polarization for the emissions
  from the beaming region with the uniform distribution of sources in
  the cases: the nonrotating dipole$-$panels $(a)$ and $(a');$
  rotating dipole$-$panels $(b)$ and $(b');$ nonrotating
  PC--current-perturbed dipole$-$panels $(c)$ and $(c');$ and rotating
  perturbed dipole$-$panels $(d)$ and $(d').$ In each panel, emissions
  are normalized with the corresponding maximum of the total
  intensity. Here, we used the parameters $\phi'=0^\circ,$ $\nu=600$
  MHz and the rest are the same as in Figure \ref{fig:Figure2}.}
 \label{fig:Figure4}
\end{figure}

By assuming an uniform distribution of sources throughout the emission
region, we computed the radiation field from the beaming region and
its polarization state, and presented in
Figure~\ref{fig:Figure4}. Since the magnitude of rotation of contour
patterns of the total intensity and the linear polarization in
$(\theta,\phi)-$ are more or less the same as that of the circular
polarization, we present only the contour patterns of circular
polarization. However, the total intensity and linear polarization
will be maximum at the beaming region center and fall considerably
towards the boundary with a slightly larger emission towards the
larger curvature region \citep[see Figure~3 in][]{KG2012a,KG2012b}.

Due to pulsar rotation, the contour patterns of the circular
polarization (panels $b,~b')$ gets rotated in $(\theta,~\phi)$--plane
with respect to those in the nonrotating dipole (panels $a,~a'),$ and
the rotation is from the trailing side to the leading side for both
the signs of $\sigma.$ Further there arises an asymmetry in the
strength of the positive and negative polarities of the circular
polarization in such a way that the negative circular gets quite
stronger.  On the other hand due to the PC--current perturbation, the
rotation direction of the contour patterns of the circular
polarization (panels $c,~c'$) becomes opposite between the positive
and negative $\sigma$ cases. That is for the positive $\sigma$ it is
opposite to that due to the effect of rotation, whereas it is in the
same direction for the negative $\sigma.$ However the magnitude of
rotation of the contour patterns due to the PC--current is smaller
than that due to the pulsar rotation. Also, due to the PC--current,
the positive polarity of the circular polarization becomes stronger
than the negative polarity for both $\pm\sigma.$ Therefore, in the
case of rotating PC--current perturbation (panels $d,~d'$), the
rotation direction of the contour pattern of the circular polarization
will be in the same direction as that in the panels $(b)$ and $(b')$
but with the lower magnitude of rotation for the positive $\sigma$
than for negative $\sigma.$ Further, due to the opposite selective
enhancement of either the positive or negative polarity of the
circular polarization by the two perturbations (rotation and
PC--current), the circular polarization becomes more or less have same
strength between the positive and negative polarities, in similarity
with the nonrotating case.

\begin{figure}
\centering
\epsscale{0.7}
\plotone{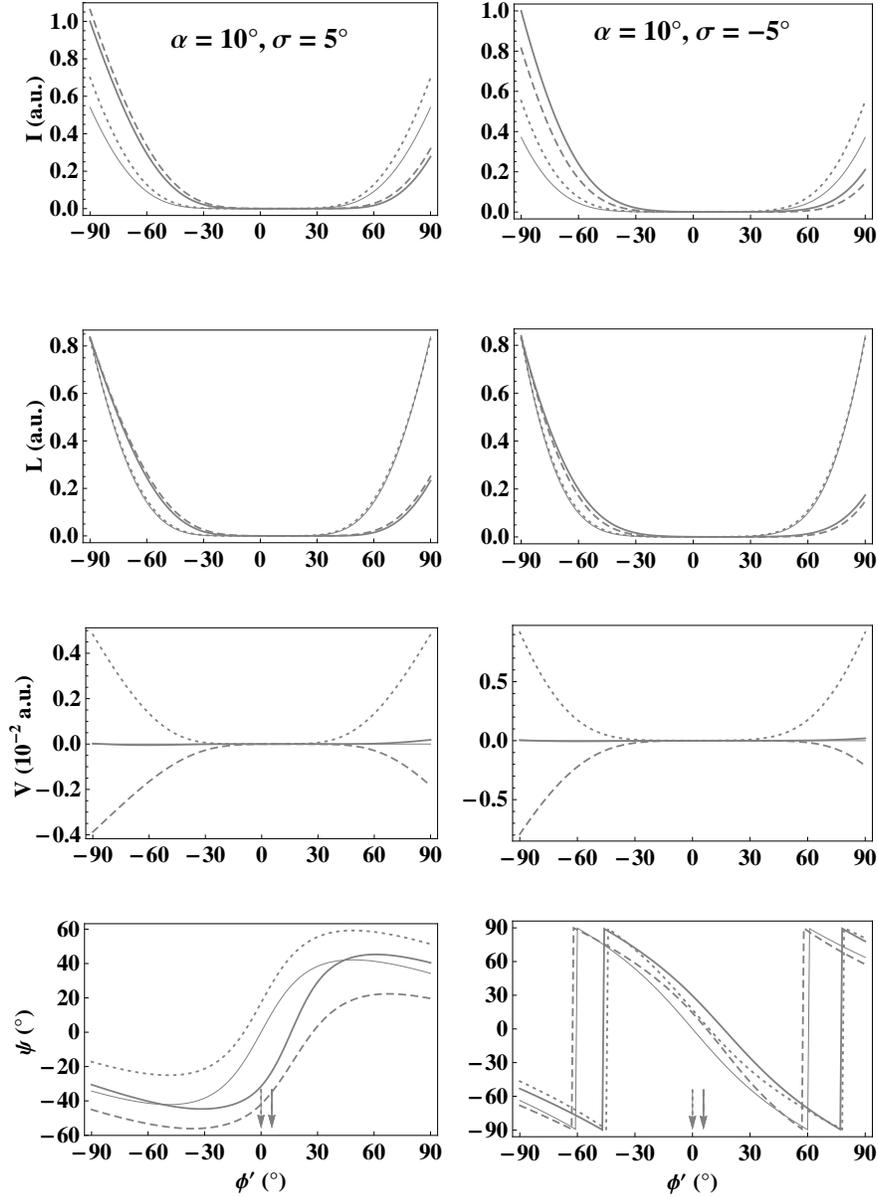}
\caption{Simulations showing the emission with uniform distribution of
  sources before and after considering the perturbations due to the
  rotation and PC--current (the lines style is the same as in Figure
  \ref{fig:Figure2}). In $I$ panels, the emissions are normalized with
  the corresponding maximum of the total intensity after combining the
  the two perturbations, whereas in $L$ and $V$ panels, they are
  normalized with the corresponding maximum of $I$ in the respective
  cases. The arrows in the $\psi$ panels (styled the same as $\psi)$
  indicate the PPA inflection points. Used $\nu=600$ MHz and the rest
  are the same as in Figure \ref{fig:Figure2}.}
 \label{fig:Figure5}
\end{figure}

Emissions from the beaming region due to different plasma bunches will
be incoherently added at the observation point if they are separated
by a space larger than the radiation wavelength. Hence, the resultant
emission that the observer receives will be the sum of the intensities
from the different plasma bunches. Using the expressions given in
\citet[from Equation 33 to Equation 36]{G2010}, we computed the
polarization state of the emitted radiation due to uniform
distribution of sources, and they are presented in Figure
\ref{fig:Figure5}. With the uniform distribution of sources, the
PC--current perturbation alone (represented by the dotted line curves)
does not affect the symmetry of the total intensity $I$ between the
leading and trailing sides of $\phi'=0^\circ$ which is similar to the
nonrotating dipole, whereas the rotation introduces an asymmetry
(represented by the dashed line curves) with the leading side becomes
stronger due to an induced larger curvature of source trajectory on
the leading side (see Figure \ref{fig:Figure3}). In the more realistic
rotating perturbed dipole, there remains an asymmetry similar to the
case of rotating dipole, however with the emission that is
significantly differs from that in the cases of the rotating dipole
and the PC--current perturbed dipole when considered separately.

Due to an incoherent addition of emissions from the different bunches
within the beaming region the magnitude of linear polarization $L$
become bit smaller than that of total intensity $I$. However, the
profile of $L$ more or less matches with the corresponding $I.$ The
net survived circular polarization $V$ in the case of rotating
perturbed dipole (represented by the thick solid line curves) becomes
very tiny because of the two perturbations (rotation and PC--current), which
selectively enhance the opposite polarities of $V$ as shown in
Figure~\ref{fig:Figure4}. The position angle is increasing in the case
of positive $\sigma$ whereas it is decreasing in the case of negative
$\sigma$ with the inflection point always shifted to trailing side.

\subsection{Emissions Due to Nonuniform distribution of Sources}
We considered a Gaussian modulation function that is given in
\citet{KG2012a,KG2012b} to model the nonuniform distribution of
sources in the emission region. We find the polarization state of the
radiation field from the nonuniform distribution of sources using the
expressions given in \citet{G2010} \citep[see equations 38
  of][]{G2010}.

\subsubsection{Emission with Azimuthal Modulation}
By considering a modulation in the magnetic azimuth with the peak at
the meridional plane, we simulated the polarization profiles affected
by the rotation and PC--current perturbations, and plotted in Figure
\ref{fig:Figure6} along with the profiles of unperturbed emissions. We
chose $r_n=0.05,$ $f_\theta=1,$ $\phi_P=0^\circ,$ $\sigma_\phi=0.1$
and the rest parameters the same as in Figure~\ref{fig:Figure5}. We
can see that due to perturbations both the emission and polarization
significantly get affected in phase and magnitude, however, the
maximum of $L$ mostly remain unaffected. Pulsar rotation causes the
intensity components to get shifted to the earlier phases with respect
to the fiducial plane, whereas the PPA inflection points to later
phases for both the signs of $\sigma.$ On the other hand PC--current
causes the intensity components shift to later phases and the PPA
inflection point to earlier phases for the positive $\sigma$ and vise
versa for the negative $\sigma.$

The net phase shift of intensity component and that of the PPA
inflection point after combining the two perturbations are found to be
$-2^\circ.16$ and $7^\circ.95,$ respectively in the case of
$\sigma=5^\circ,$ whereas they are found to be $-6^\circ.69$ and
$8^\circ.90,$ respectively in the case of $\sigma=-5^\circ.$ On the
other hand, the sum of the phase shifts of the intensity components
caused separately by the rotation and PC--current are found to be
$-2^\circ.10$ and $-6^\circ.87$ respectively in the cases of
$\sigma=\pm5^\circ,$ whereas that of the PPA inflection point are
found to be $6^\circ.44$ and $9^\circ.59,$ respectively. 
Therefore, the absolute relative difference between the phase shift of
intensity component that resulted when the two perturbations taken
together and that due to the sum of two separate perturbations, is
about $<3\%$ in both the cases of $\sigma,$ whereas that for the PPA
inflection point is found to be about $19\%$ in the case of
$\sigma=5^\circ,$ and $8\%$ in the case of $\sigma=-5^\circ.$
Note that the net phase shift of the intensity
component becomes slightly larger than that of the net modulation $f$
(which is about $-1^\circ.92$ and $-5^\circ.71,$ respectively in the
cases of $\sigma=\pm5^\circ$) due to an induced asymmetry in the net
radius of curvature about the peak location of modulation.

Although circular polarization of opposite polarities from the
background unmodulated emission roughly cancels out, a net circular
however with an asymmetry between the opposite polarities survives in
the presence of modulation.  This is because of an asymmetry in the
magnitude of rotation of the emission pattern with respect to rotation
phase in such a way that larger rotation magnitude towards the inner
rotation phases as compared to that on outer phases \citep{KG2012a}.
Hence, it results in the selective enhancement of the leading side
circular over the trailing side circular.  However, the asymmetry
between the opposite polarities of the circular polarization becomes
smaller for the positive $\sigma$ than that for the negative $\sigma.$
This is due to an opposite behavior of PC--current in introducing an
asymmetry between the opposite polarities of $V$ between the
$\pm\sigma$ wherein it selectively enhances the trailing positive
circular for $+\sigma$ whereas the leading positive circular for
$-\sigma.$ On the other hand pulsar rotation causes the selective
enhancement of the leading polarity of $V,$ i.e., negative circular
for $+\sigma$ and positive circular for $-\sigma.$

To assess whether we can estimate the net phase shift of intensity
components by adding the phase shift due to each perturbation computed
separately, we chose emission altitude $r_n=0.1,$ and two cases of
modulation: narrower $(\sigma_\phi=0.1)$ and broader
$(\sigma_\phi=0.3).$ The simulated pulses are presented in
Figure~\ref{fig:Figure7}. The net phase shift of the component after
combining the two perturbations in the cases of $\sigma_\phi=0.1$ and
$0.3$ are found to be $-4^\circ.40$ and $-8^\circ.33,$ respectively,
whereas the PPA inflection point phase shifts are found to be
$15^\circ.70$ and $16^\circ.23,$ respectively.  On the other hand the
sum of the phase shifts of the intensity components caused due to the
two perturbations considered separately are found to be $-4^\circ.09$
and $-5^\circ.77$ respectively wherein the relative difference are
about $7\%$ and $30\%$ respectively.  For PPA inflection point the
corresponding phase shifts are found to be $14^\circ.52$ and
$16^\circ.52,$ respectively, wherein the absolute relative differences
are about $7.5\%$ and $2\%,$ respectively.  Hence the actual phase
shifts of intensity components and PPA inflection point will be
significantly different from those obtained when the two perturbations
dealt separately.

In the case of $\sigma_\phi=0.1,$ the asymmetry in the strength of
circular polarization between the opposite polarities become larger as
compared to that in the lower altitude emission (see Figure
\ref{fig:Figure6}). This is due to an increased asymmetry in the
magnitude of rotation of beaming region emission pattern between the
inner and outer phases wherein the small amount of rotation occurs at
the outer phases. In the case of $\sigma_\phi=0.3,$ only the leading
negative polarity of the circular polarization survives due to
aforementioned reasons but this time it is enhanced more due to
broader modulation and hence broader pulse width. Hence it results in
the ``symmetric''-type circular polarization.

As a case of modulation which lies symmetrically on either side of the
meridional plane, we choose the modulation peaks located at
$\phi_P=\pm30^\circ$ for $\sigma=5^\circ$ and
$\phi_P=180^\circ\pm15^\circ$ for $\sigma=-5^\circ,$ and the simulated
profiles are given in Figure \ref{fig:Figure8}.  Note that only the
combined case of rotation and PC--current perturbations is given in
all panels.  Although an observer encounters mostly the same plasma
density (modulation strength) between the leading and trailing sides,
the net modulated intensity component on the trailing side becomes
weaker than that on the leading side due to the induced larger
curvature of source trajectory on the leading side. Further the
leading side component becomes broader than the trailing one due to an
induced asymmetry in the sight line encountered modulation and that in
the gradient of radius of curvature. The behaviors of linear and
circular polarization are same as in the Figure \ref{fig:Figure6}
except the enhancement of trailing part of the circular polarization
in the trailing side components.

\subsubsection{Emission with Polar Modulation}
\begin{figure}
\centering
\epsscale{0.6}
\plotone{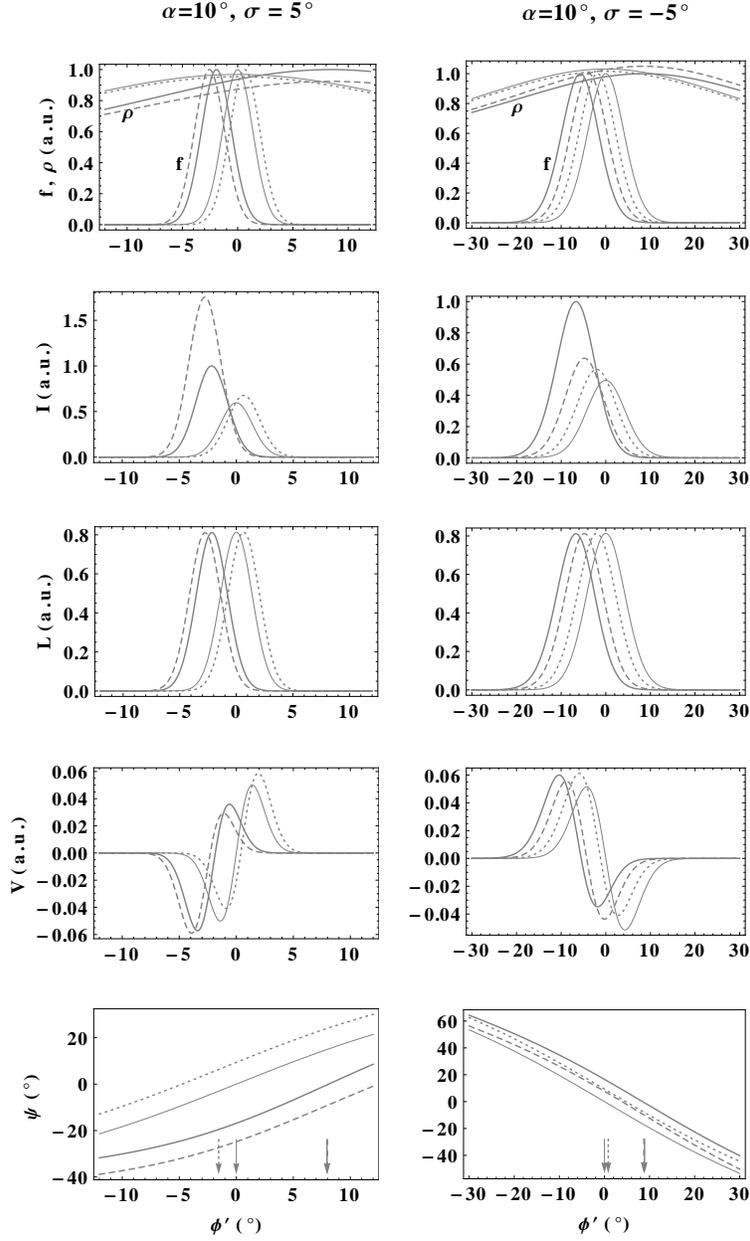}
\caption{Simulated pulse profiles with modulation in the azimuthal
  direction. The total intensity $I$ is normalized with the maximum of
  $I$ of the combined case of rotation and PC--current, whereas $L$ and
  $V$ are normalized with the corresponding maximum of $I$ in the
  respective cases. We used $r_n=0.05,$ $f_\theta=1,$
  $\phi_P=0^\circ,$ and $\sigma_\phi=0.1$. The rest parameters are the
  same as in Figure \ref{fig:Figure5}.}
 \label{fig:Figure6}
\end{figure}

\begin{figure}
\centering
\epsscale{0.7}
\plotone{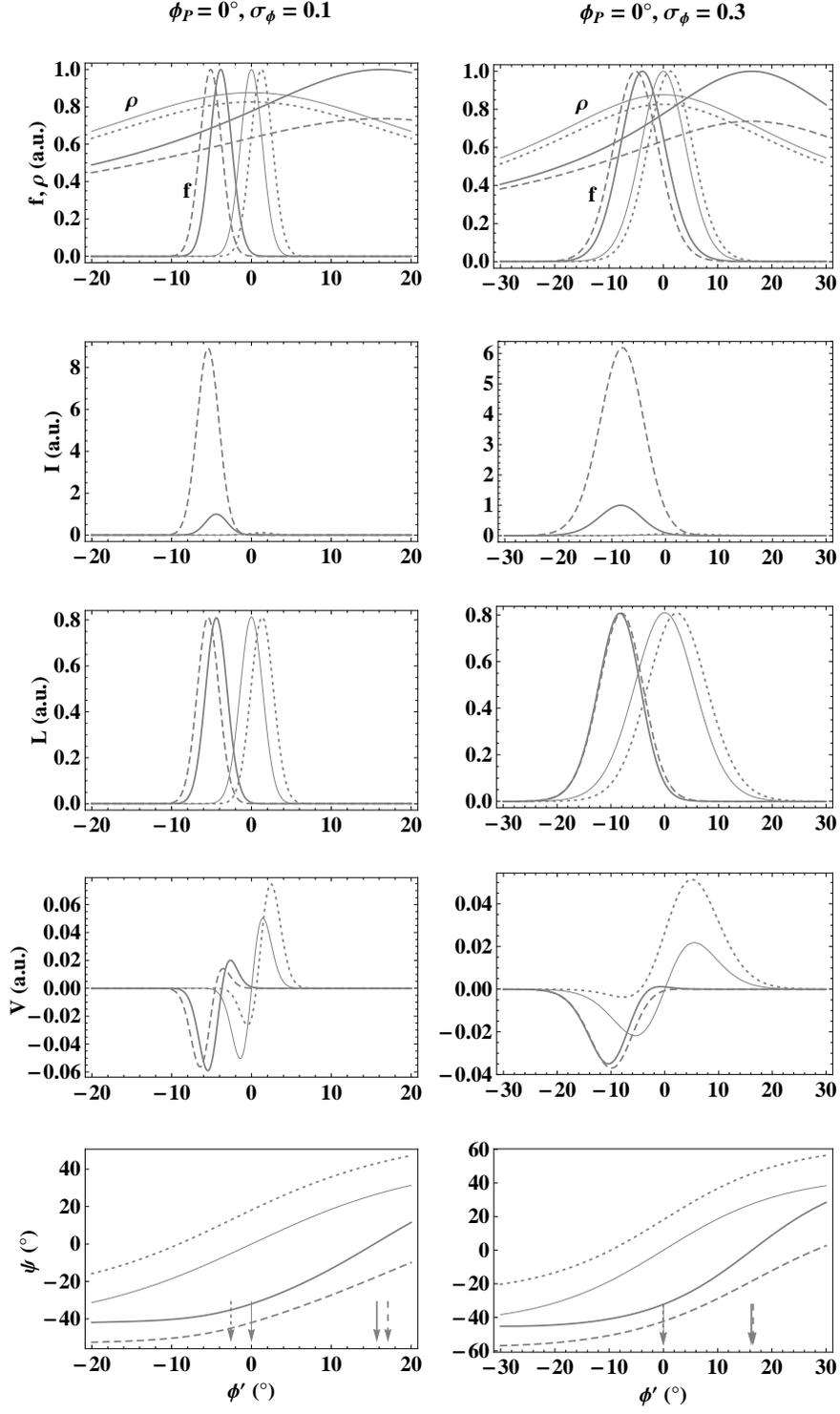}
\caption{Same as Figure \ref{fig:Figure6} except $r_n=0.1,$
  $\sigma=5^\circ$ and two types of modulation: narrower
  $(\sigma_\phi=0.1)$ and broader $(\sigma_\phi=0.3)$ Gaussians are
  considered. }
 \label{fig:Figure7}
\end{figure}

\begin{figure}
\centering
\epsscale{1.0}
\plotone{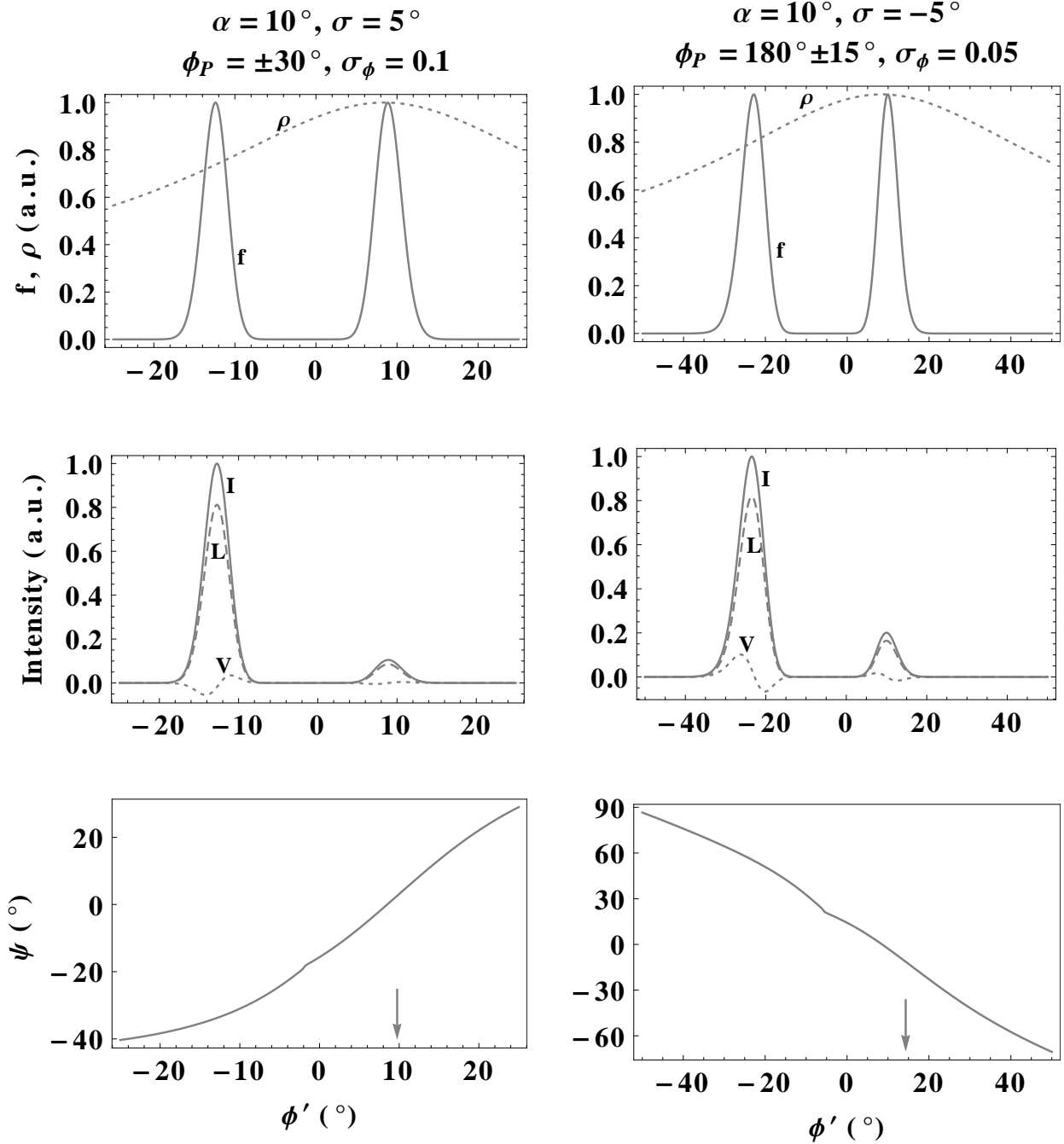}
\caption{Same as Figure \ref{fig:Figure6} but with two symmetrically
  located Gaussian modulations with respect to magnetic meridian plane
  and only the combined case of rotation and PC--current perturbations
  is presented. In the upper panels: $\rho-$dotted line curves,
  $f-$solid line curves; in the intensity panels: $I-$ solid line
  curves, $L-$dashed line curves, $V-$ dotted line curves.}
 \label{fig:Figure8}
\end{figure}

\begin{figure}
\centering
\epsscale{0.65}
\plotone{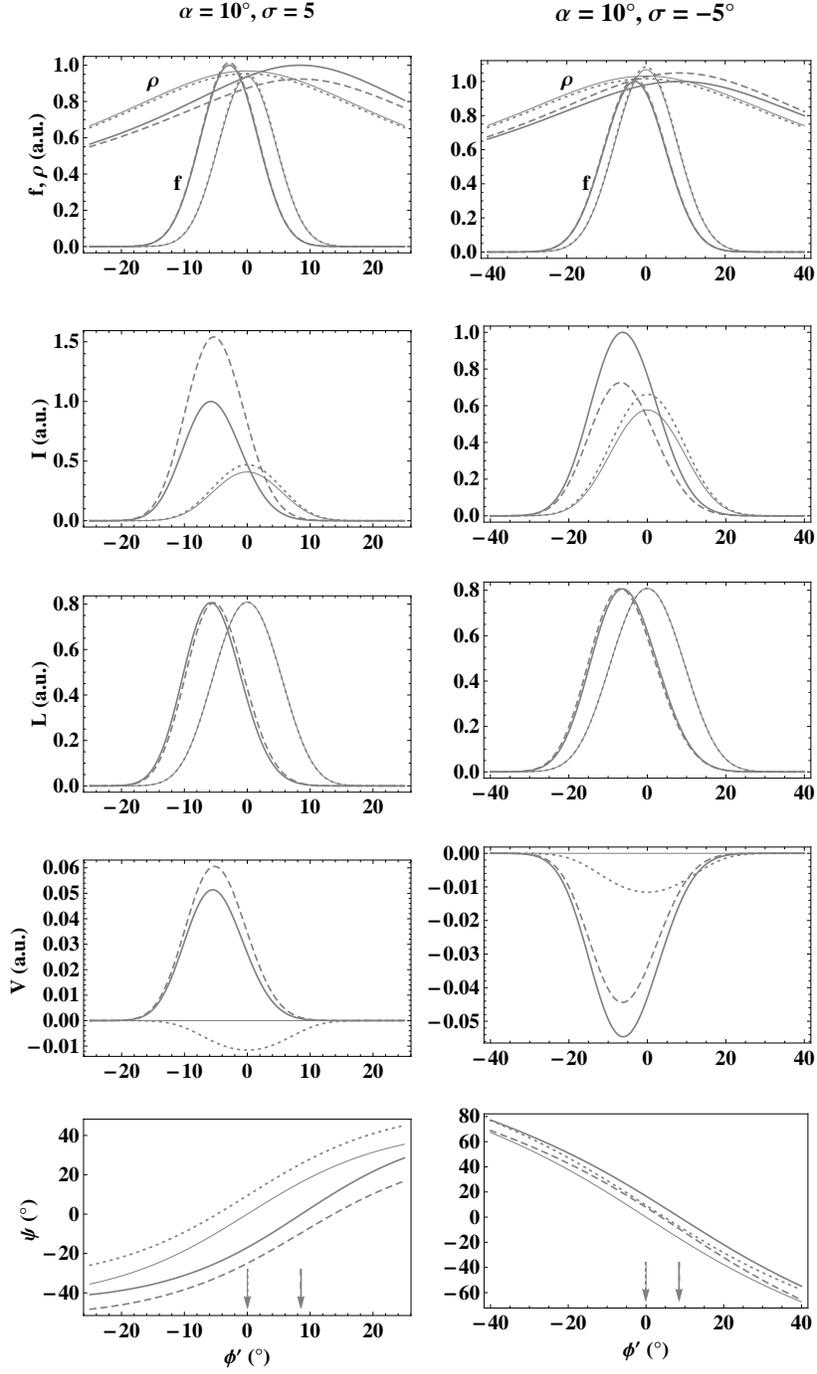}
\caption{Same as Figure \ref{fig:Figure6} but with the nonuniform
  distribution of sources in the polar direction and $f_\phi=1$,
  $\theta_P=2^\circ,$ and $\sigma_\theta=0.01$.}
 \label{fig:Figure9}
\end{figure}

By considering a modulation with plasma density gradient in the polar
direction we present the effects of rotation and PC--current
perturbation on pulsar emission and polarization.  In Figure
\ref{fig:Figure9}, we present the simulation of hallow cone emissions
surrounding the magnetic axis with modulation peak at
$\theta_P=2^\circ.$ Since the minimum of $\theta_0$ is about
$\sim(2/3)\sigma$ which is about $3^\circ.3$ for $\sigma=\pm5^\circ,$
observer's sight line cuts the hollow cone emission only once in each
pulsar rotation, and hence it results in a single component profile.
For the sake of comparison, the cases of before and after
consideration of the perturbations are all shown in the Figure
\ref{fig:Figure9}, and the parameters normalization and the line
representations are the same as in Figure \ref{fig:Figure6}.  Note
that, unlike the case of azimuthally modulated emissions (see Figure
\ref{fig:Figure6}), the maximum of modulation strength that the
observer encounters in the lab frame is not same in all the cases as
the minimum of $\theta_0$ slightly get affected due to the
perturbations. But, similar to the case of Figure \ref{fig:Figure6},
the emissions get affected significantly before and after considering
the perturbations due to the induced differences in $\rho$ at the peak
locations of $f.$ The net phase shifts of a component after combining
the two perturbations in the cases of $\sigma=\pm5^\circ$ are found to
be $-5^\circ.79$ and $-6^\circ.29,$ respectively.  On the other hand
the net phase shift obtained by adding the phase shifts due to the
rotation and PC--current when considered separately in the cases of
$\sigma=\pm5^\circ$ are found to be $-5^\circ.32$ and $-6^\circ.83,$
respectively. Hence the relative differences are about $8\%$ and
$8.5\%,$ respectively with respect to the combined case of the
perturbations. Note that this relative differences in the intensity
phase shifts become larger at higher altitude $(\sim r_n=0.1).$

The maximum of normalized linear polarization 
$L$ is more or less the same in all the cases, similar to Figure
\ref{fig:Figure6}.  Although the position angle significantly gets
affected after combining the two perturbations, its inflection point
lies at roughly the same phase as that due to the rotation alone.  The
circular polarization $V$ becomes symmetric type with the opposite
polarities survive due to an opposite direction of rotation of
emission pattern in the beaming region.  The net circular polarization
after combining the two perturbations will also be symmetric type with
the survival of polarity as that due to the pulsar rotation alone.

\begin{figure}
\centering
\epsscale{1.0}
\plotone{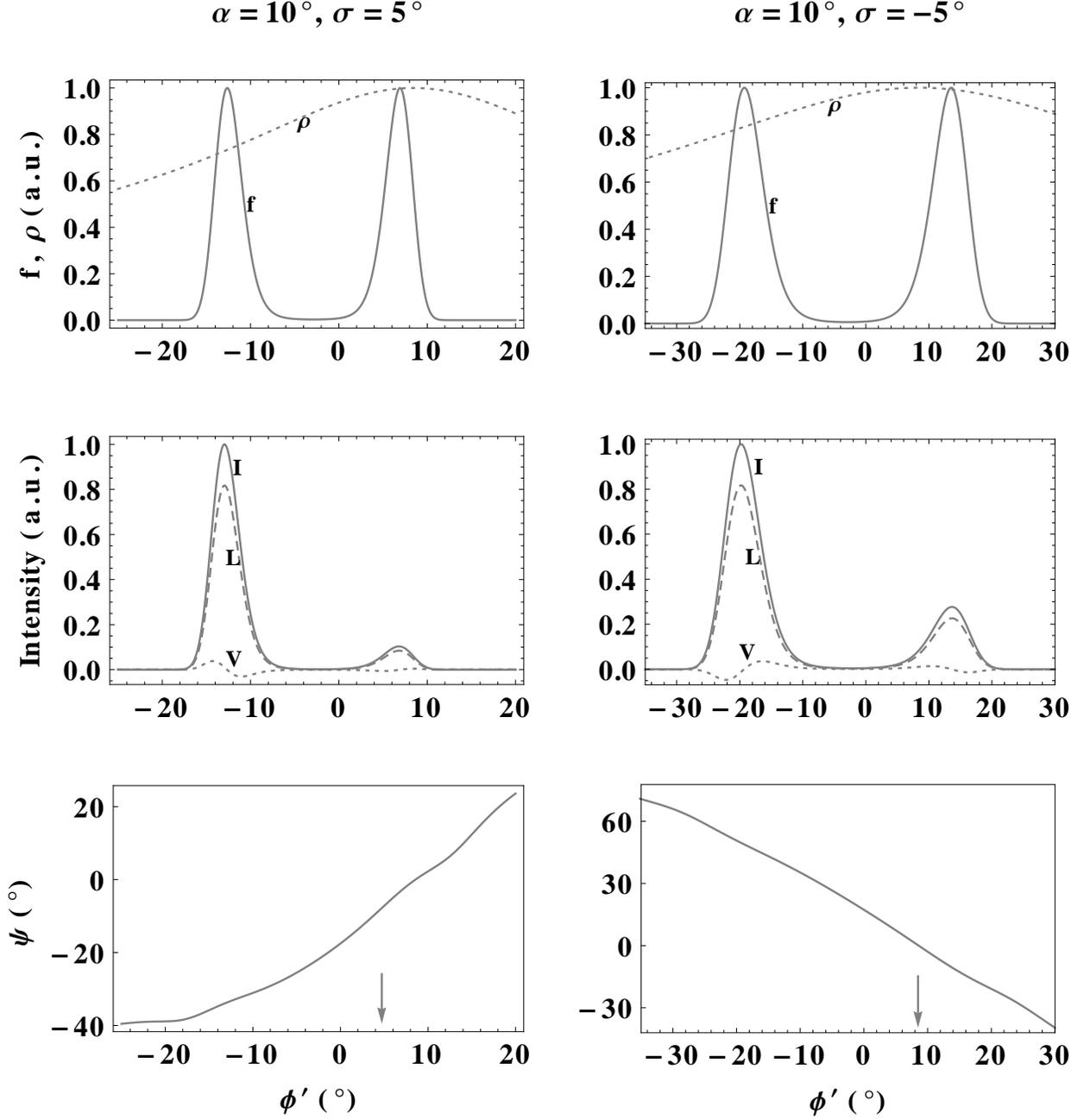}
\caption{Same as Figure \ref{fig:Figure9} with the combined effect of
  rotation and PC--current (the line style is the same as in Figure
  \ref{fig:Figure8}) except with the parameters $\theta_P=3^\circ.6$
  and $\sigma_\theta=0.002.$}
 \label{fig:Figure10}
\end{figure}

By considering a hallow cone modulation with peak at
$\theta_P=3^\circ.6,$ we have analyzed a case where the sight line
cuts the hallow cone emission twice (see Figure \ref{fig:Figure10}).
Similar to Figure \ref{fig:Figure8}, the trailing side intensity
component become weaker as well as narrower than that on the leading
side, due to the induced asymmetry in the curvature of source
trajectories between the two sides.  Since the sight line crosses the
central maximum of the hallow cone on both the leading and trailing
sides, there is a change over of selective enhancement of emission
over the part of the beaming region with smaller values of $\theta$ to
that over the larger values $\theta$ on the leading side, and vice
versa on the trailing side.  Hence, there results an antisymmetric
circular polarization with the sign reversal from positive to negative
over in leading side component and vice versa on the trailing side for
the case of positive $\sigma.$ On the other hand, it is vice versa for
the case of negative $\sigma$ due to the opposite direction of
rotation of the emission pattern of the beaming region.

\subsubsection{Emission with Modulation in both Azimuthal and Polar
  directions } 
\begin{figure}
\centering
\epsscale{0.9}
\plotone{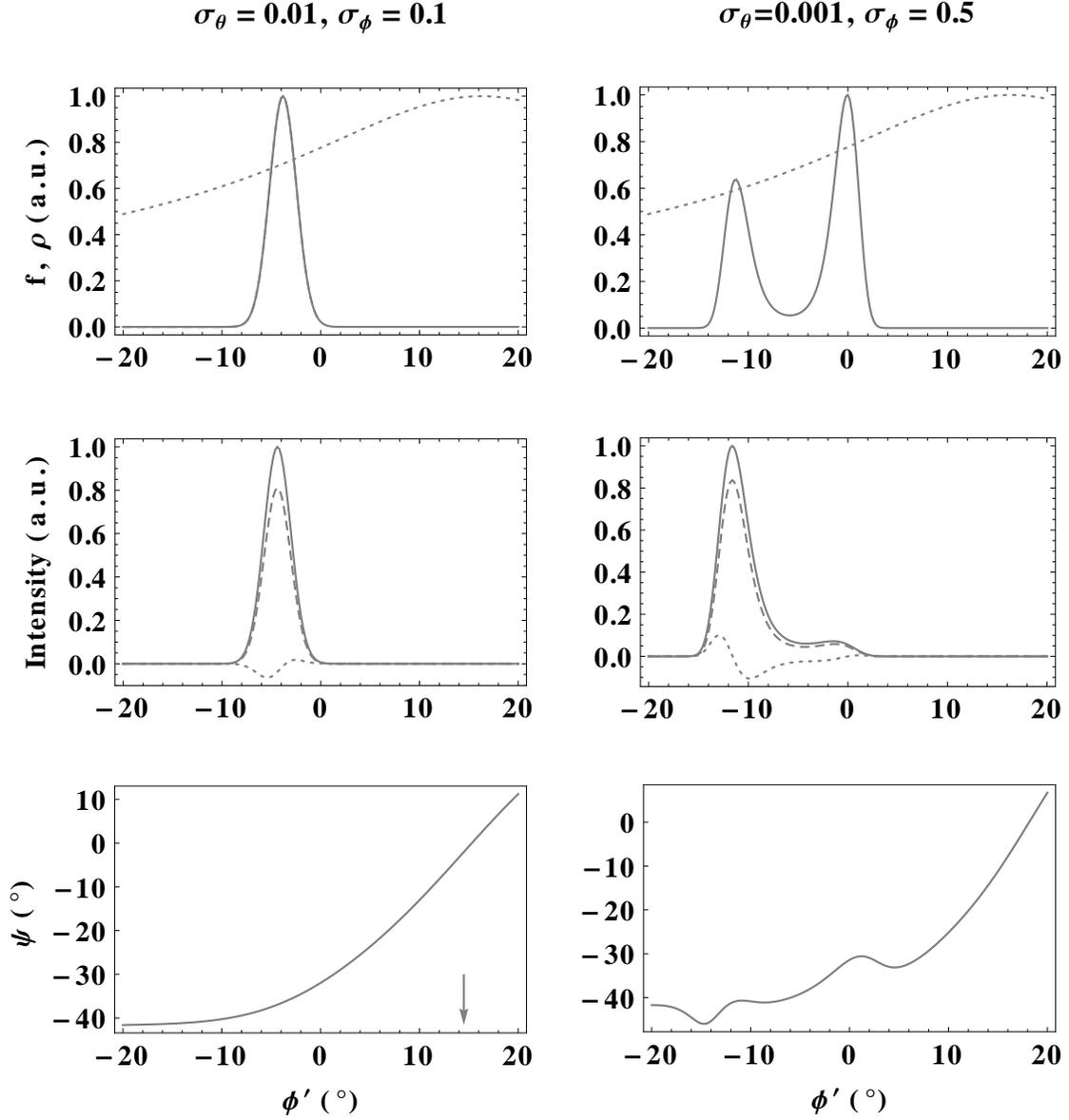}
\caption{Simulated pulse profiles with the nonuniform distribution of
  sources in both the polar and azimuthal directions, after considering
  the perturbations by the rotation and PC--current. The line style is
  the same as in Figure \ref{fig:Figure8}.  We chose $\sigma=5^\circ,$
  $r_n=0.1,$ $f_0=1,$ $\theta_p=3^\circ.4,$ $\phi_P=0^\circ,$ and the
  rest parameters are the same as in Figure \ref{fig:Figure5}.}
 \label{fig:Figure11}
\end{figure}

  The radiation sources may be nonuniformly distributed
in both the polar and azimuthal directions in the pulsar
magnetosphere. The extreme cases, wherein the modulation is
predominant in azimuthal or polar directions are already discussed in
Sections $3.2.1.$ and $3.2.2.,$ respectively.  In this section we
present a few more cases where the modulation exists in both the polar
and azimuthal directions.  In Figure \ref{fig:Figure11}, we considered
the cases $\sigma_\theta=0.01$ and $\sigma_\phi=0.1$ wherein the
modulation is effectively dominating in the azimuthal direction over
that in the polar direction, and $\sigma_\theta=0.001$ and
$\sigma_\phi=0.5$ wherein the modulation becomes effective in
$\theta.$ Note that in the case of $\sigma_\theta=0.01$ and
$\sigma_\phi=0.1,$ even though $\sigma_\theta<\sigma_\phi,$ the
modulation becomes effective in $\phi$ coordinate than in $\theta$ due
to the much larger coverage of $\phi$ compared to that of $\theta,$
see for e.g., Figures \ref{fig:Figure3} and \ref{fig:Figure4}.  In the
case of $\sigma_\theta=0.01$ and $\sigma_\phi=0.1,$ the emission and
polarization properties are similar to the case of $\alpha=10^\circ$
and $\sigma=5^\circ$ of Figure \ref{fig:Figure6} with an antisymmetric
circular polarization wherein the sign reversal is from the negative
polarity to the positive.  On the other hand even though the single
modulation is considered in the case of $\sigma_\theta=0.001$ and
$\sigma_\phi=0.5$ but due to the viewing geometry, much elongation of
the modulation in the azimuthal direction and squeezing into a narrow
cone, the modulation encountered by the sight line results in a
blended two components like structure.  It further results into a two
components like structure in intensity profile, however, with much
weaker trailing side due to the larger $\rho.$ Although circular
polarization is still antisymmetric type, the sign reversal becomes
opposite to the case of $\sigma_\theta=0.01$ and $\sigma_\phi=0.1,$
i.e., from positive to negative, which is similar to the case of
$\alpha=10^\circ$ and $\sigma=5^\circ$ in Figure \ref{fig:Figure9}.
Hence in the presence of perturbations, it is hard to see the
correlation between the sign reversal of circular polarization and the
PPA swing as both the types of circular sign reversal seems to be
associated with the increasing PPA swing.  Note that, the inflection
point of PPA swing is not derived in the case of $\sigma_\theta=0.001$
and $\sigma_\phi=0.5$ due to the difficulty in finding it because of
the kinky nature in the PPA swing.

\begin{figure}
\centering
\epsscale{1.0}
\plotone{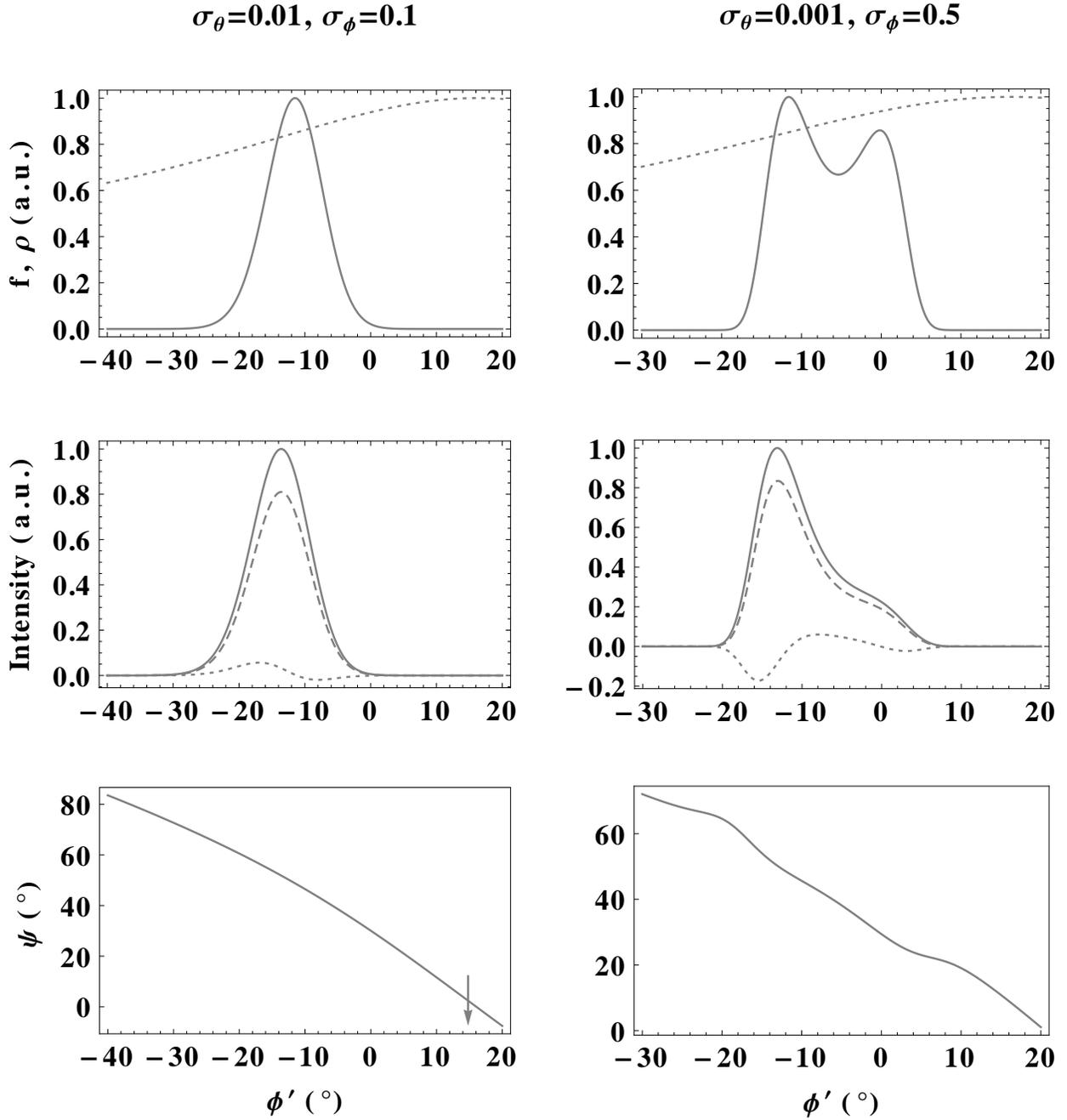}
\caption{Same as Figure \ref{fig:Figure11} except with
  $\sigma=-5^\circ$ and $\phi_P=180^\circ.$ }
 \label{fig:Figure12}
\end{figure}

By using $\sigma=-5^\circ,$ $\phi_P=180^\circ$ and the rest parameters
the same as in Figure \ref{fig:Figure11}, we computed the polarization
profiles and plotted in Figure \ref{fig:Figure12}.  In both cases of
modulation, the profiles are similar to those in Figure
\ref{fig:Figure11} except with the opposite sign reversal of the
circular polarization in the respective cases and decreasing PPA
swing.  Also, similar to Figure \ref{fig:Figure11}, due to
perturbations no correlation is found between the sign reversal of the
circular polarization and the PPA swing, as both the sign reversal of
the circular polarization, i.e., from positive to negative or vice
versa, are associated with the decreasing PPA.

\begin{figure}
\centering
\epsscale{1.0}
\plotone{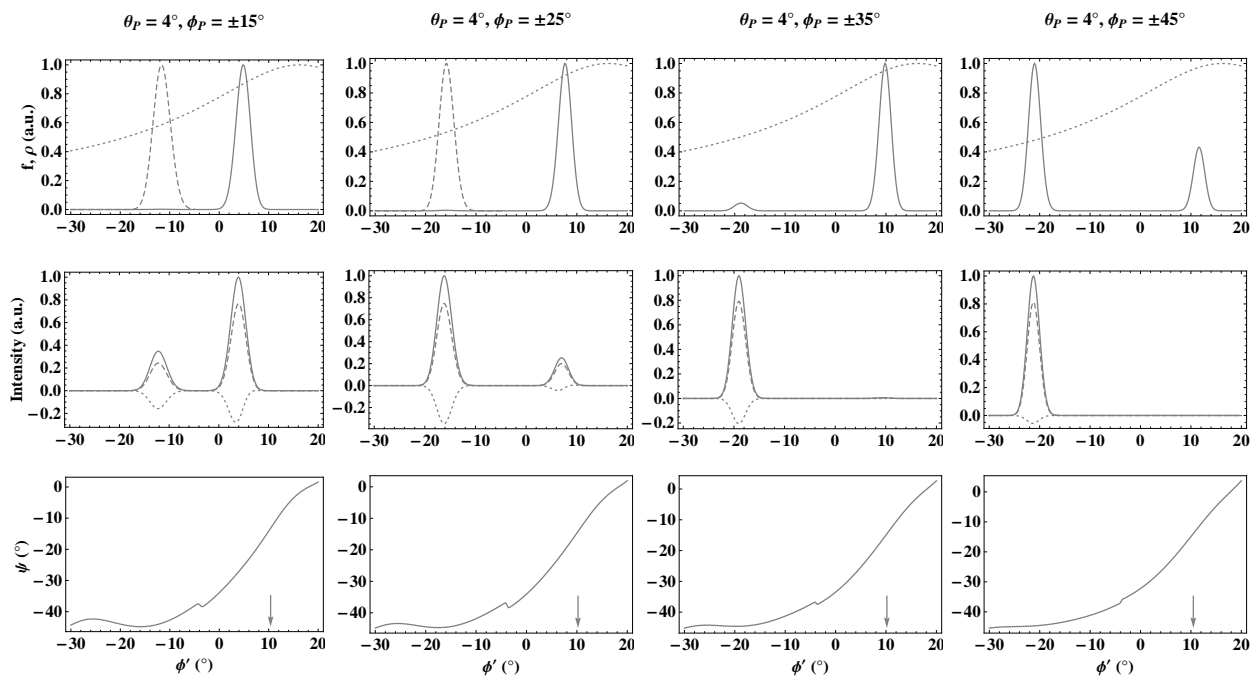}
\caption{Same as Figure \ref{fig:Figure11} but with two Gaussian
  modulations symmetrically located in a given conal ring at various
  azimuth. For simulation we used $\sigma_\theta=0.003,$
  $\sigma_\phi=0.1$ and the rest parameters are the same as in Figure
  \ref{fig:Figure11}. The dashed line curves in the first two columns
  of the upper panels represent an amplified $f$ to unity to identify
  with the corresponding intensity component.}
 \label{fig:Figure13}
\end{figure}

In Figure \ref{fig:Figure13}, we presented the simulations for
different cases of two Gaussian modulations symmetrically located in a
given ring centered on the magnetic axis.  Although, the asymmetry in
$\rho$ between the leading and trailing sides is same in all the cases
with much larger curvature on the leading side, there is a diverse
asymmetry in the strength of the intensity component between the
leading and trailing sides.  This is due to an induced asymmetry in
the strength of modulation that the inertial observer encounters
between the leading and trailing sides.  In the case of
$\theta_P=4^\circ$ and $\phi_P=\pm15^\circ,$ observer encounters a
much weaker modulation on the leading side than that on the trailing
side which overcomes the influence of asymmetric $\rho.$ Hence, it
results in a stronger trailing side component than that on the leading
side.  Whereas in the case $\theta_P=4^\circ$ and
$\phi_P=\pm25^\circ,$ observer encountered asymmetry in the modulation
strength between the leading and trailing sides becomes smaller
compared to that in the case $\theta_P=4^\circ$ and
$\phi_P=\pm15^\circ.$ Hence the leading side intensity component
becomes stronger than the trailing one as the influence of asymmetry
in $\rho$ becomes much important than that in $f.$ In case of
$\theta_P=4^\circ$ and $\phi_P=\pm35^\circ,$ the asymmetry in the
modulation strength between the leading and trailing sides becomes
even smaller, and hence resulted in mostly in leading side component. In
case of $\theta_P=4^\circ$ and $\phi_P=\pm45^\circ,$ observer
encounters higher modulation strength on the leading side than that on
trailing side, and hence resulted in a single leading side component
or a partial cone.

In all the cases, circular polarization is found to be symmetric type
over both the leading and trailing side components due the selective
enhancement of negative polarity caused by the rotation and
PC--current perturbation.  Note that for example in the case of
$\theta_P=4^\circ$ and $\phi_P=\pm15^\circ,$ the emissions from the
region, which is closer to the magnetic axis, are selectively
enhanced, and hence a selective enhancement of inner circular occurs
(see for e.g., the case $\alpha=10^\circ$ and $\sigma=5^\circ$ in
Figure \ref{fig:Figure10}). The kinks are
introduced into the PPA swing due to the combined effect of
modulation and perturbation caused by rotation and PC--current.

Similar to Figure \ref{fig:Figure13}, we selected Gaussian modulations
which are symmetrically located on either sides of the magnetic
meridian plane in a given cone, and the simulated profiles are shown
in Figure \ref{fig:Figure14} for $\sigma=-5^\circ.$ Similar to the
cases in Figure \ref{fig:Figure13}, the modulation strength that the
observer encounters can be quite different between the leading and
trailing sides in addition to an asymmetric $\rho.$ Hence it leads to a larger asymmetry
in the strength of components between the two sides.  In the case
$\phi_P=180^\circ\pm10^\circ$ observer misses out the trailing side
modulation and hence only the leading side component shows up, whereas
in the case $\phi_P=180^\circ\pm40^\circ$ it becomes vice versa.  For
third column, we used the combined modulation that are used in the
first two columns (four Gaussian modulations). Even though
observer encounters a weaker modulation on the leading side, due to
larger curvature on the leading side the component on the leading side
becomes stronger than the corresponding one on trailing side.  But in
the case of last column panels, observer encounters even weaker
modulation on the leading side than on the trailing side as compared
to the case of third column.  Hence the resultant leading side
component becomes weaker than the corresponding trailing side one.  In
all the cases, similar to Figure \ref{fig:Figure13}, due to the
selective enhancement, the circular polarization becomes positive
symmetric type over both the leading and trailing sides.
 
\section{DISCUSSION}
The poloidal PC--current perturbs the underlying dipole
field by inducing a toroidal magnetic field \citep{HA2001, KG2012b}
whereas the pulsar rotation causes the bending of trajectory of the
field line constrained plasma in the direction of pulsar rotation
\citep[e.g.,][]{BCW1991,G2005,TG2007,D2008,DWD2010,Tetal2010,KG2012a,Wetal2012}.  They
significantly affect the phase locations of the intensity components
as well as the PPA inflection point by introducing an asymmetric phase
shifts between them.  Rotation (aberration) shifts intensity to the
leading side and the PPA inflection point to the trailing side
irrespective of sign of $\sigma$ \citep[for
e.g.,][]{BCW1991,G2005,TG2007,D2008,DWD2010,Tetal2010,KG2012a,Wetal2012}.  On the
other hand PC--current along with modulation can shift the intensity
to trailing side and the PPA inflection point to the leading side for
the positive $\sigma,$ and vice versa for negative $\sigma$
\citep{KG2012b}.  The influences of the rotation and PC--current adds up
for the negative $\sigma$ but cancel each other when $\sigma$ is
positive.  Note that the opposite behavior of rotation and PC--current
in the case of positive $\sigma$ is due to the opposite directions of
induced curvature caused by the two perturbations.

Although the effects of rotation and PC--current perturbation can be
understand qualitatively when they are combined together, their
quantitative estimate becomes impossible when the two effects are
considered separately as shown in sections $2$ and $3.$ However, since
the influence of rotation is mostly larger than that of the
PC--current, in the combined cases mostly the effects of rotation
prevail with quite different magnitudes.  For example, the phase delay
between the PPA inflection point and the central component of pulse
profile becomes much smaller for the positive $\sigma$ as compared to
that for the negative $\sigma$ (see Figure \ref{fig:Figure6}).
Hence the above said phase delays greatly depend on the geometric
parameters which has not been reported in the literature.

\begin{figure}
\centering
\epsscale{1.0}
\plotone{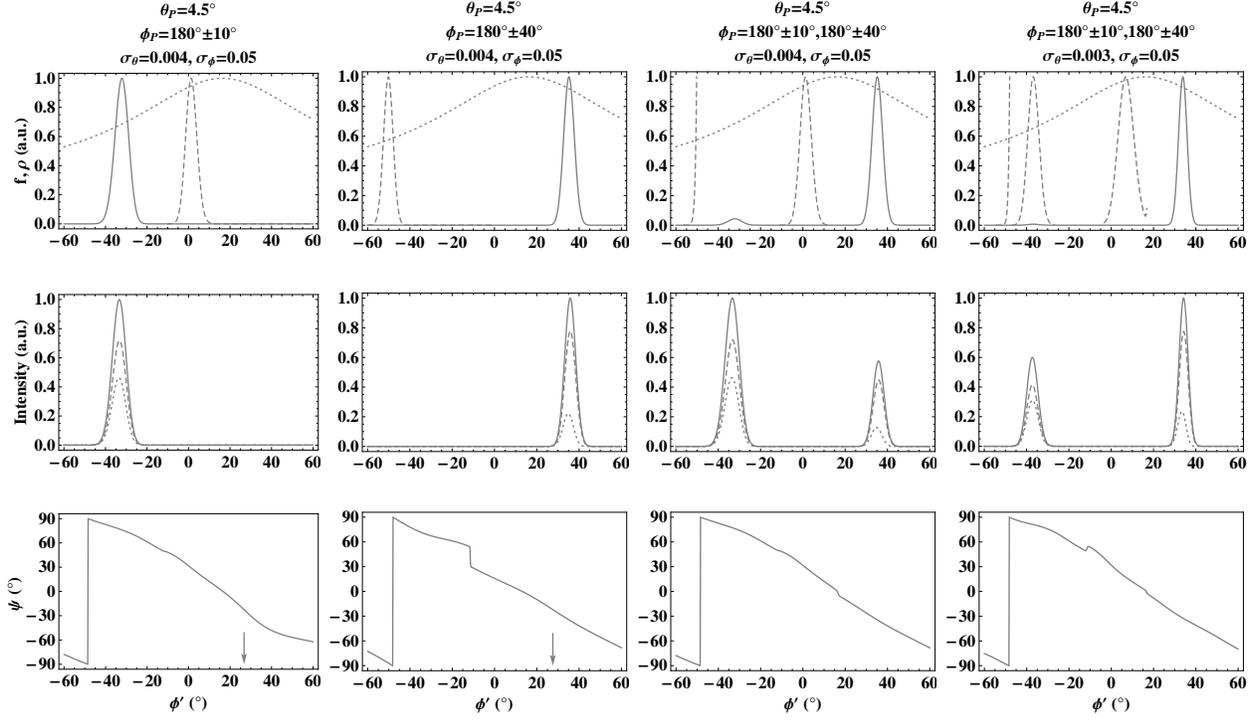}
\caption{Same as Figure \ref{fig:Figure13} except with
  $\sigma=-5^\circ$ and two Gaussian modulations considered in the
  first two column panels whereas four Gaussian modulations in the
  third and last column panels with different modulation parameters.}
 \label{fig:Figure14}
\end{figure}

Even though the curvature of source trajectory on the leading side
becomes larger than that on the trailing side due to the perturbation
caused by the rotation and PC--current considered together, the net
modulated intensity can become stronger either on the leading side or
on the trailing side (see Figures \ref{fig:Figure13} and
\ref{fig:Figure14}) depending upon the modulation location and the
viewing geometry. Statistically, although there is an observational
support for the leading side component dominating over the trailing
one, the other contrary cases are also reported \citep{LM1988}.  Hence
our model provides a much plausible explanation for the usual
asymmetry between the leading and trailing side
components. \citet{KG2012b}, by considering the PC--current alone,
also have predicted above behavior in the absence of strong rotation
effect. However by considering effect of pulsar rotation on the field
line constrained plasma alone, \cite{BCW1991} have predicted that the leading side intensity dominate over the trailing side and later \citet{TG2007}, \citet{D2008},
\citet{DWD2010}, \citet{Tetal2010}, \citet{KG2012a}, \citet{Wetal2012}
have arrived at the same conclusion. The
``partial cone'' pulsars, which show either missing or much suppressed
side in their conal double component profiles \citep{LM1988} can be
result due to the rotation and PC--current perturbation (see Figures
\ref{fig:Figure13} and \ref{fig:Figure14}).

Further the rotation and PC-current perturbation introduce an
asymmetry into the width of components between the leading and
trailing sides in such a way that the leading side components become
broader than the corresponding trailing ones, and it has an
observational support too \citep{AG2002}.  This is due to an induced
asymmetry in the steepness of radius of curvature with respect to the
rotation phase wherein the trailing side $\rho$ becomes more steeper
than that on the leading side.  \citet{KG2012a} and \citet{KG2012b} have
also predicted the same behavior by considering the effects of
rotation and PC--current separately.

The rotation and PC--current significantly introduce an asymmetry into
the strengths of opposite polarities of the circular polarization.
When they are considered separately, they selectively enhance either
the negative or positive polarities.  Further the two effects cause
the emission pattern to rotate in $(\theta,~\phi)-$plane in the
opposite directions for positive $\sigma$ and in same direction for
negative $\sigma.$ Hence it results in smaller rotation of the
emission pattern in $(\theta,~\phi)-$plane for the positive $\sigma$
case than that for the negative $\sigma.$ However due to the rotation
of the emission pattern, the symmetric type circular polarization
becomes evident in addition to the more common antisymmetric type with
an usual asymmetry between the opposite polarities in presence of
nonuniform distribution of sources \citep[see
  also][]{KG2012a,KG2012b}.
Due to perturbations we find the sign reversal of circular
polarization from negative to positive near the center of pulse
profile can be associated with either the increasing or decreasing PPA
swing, and similarly the vice versa (see Figures \ref{fig:Figure11}
and \ref{fig:Figure12}), and these deductions are found to be in accordance with
\citet{Hanetal1998} and \citet{YH2006} observational findings. 

Our simulations (see Figures \ref{fig:Figure13} and
\ref{fig:Figure14}) also confirm \citet{Hanetal1998} and
\citet{YH2006} findings that the negative polarity of the circular
polarization is associated with increasing PPA and vice versa for
positive polarity on both the sides of conal-double pulsars
profiles. This is possible due to combined rotation and PC-current
perturbation under specific conditions that when the modulation is
more effective in the polar direction than in azimuthal direction.
However \citet{KG2012a} have claimed that such correlation can exist
provided in addition to the effective modulation in the polar
direction, their location must be asymmetric in a given conal ring
centered on the magnetic axis.

Some pulsars particularly millisecond pulsars show the polarization
angle behavior that deviates from the standard `S' curve
\citep{Xetal1998}. We find that the `kinky' type distortions in PPA
profile are due to the superposition of modulated incoherent emissions
over the regions which are larger than the wavelength of radiation in
presence of strong rotation and PC-current perturbations. Note that
\citet{G2010} and \citet{KG2012a} also have obtained the similar
results. However, the results of \citet{G2010} were in the absence of any
perturbation and hence the PPA distortions from the standard `S' curve
are smaller. \citet{MS2004} have speculated that if the radio emission
has a varying emission height across the pulse profile then PPA swing
will be nonuniform. Also, in the literature the kinky type origin has
been attributed to multi polar magnetic field in the radio emission
region \citep{Metal2000} and to the return currents in the pulsar
magnetosphere \citep{RK2003}.

Note that we have performed modeling of coherent pulsar radio emission in terms of single particle curvature radiation. Although to the first order it is not a bad assumption, in reality the efficiency of coherence process may depend on many factors and hence there may be an induced asymmetries in the pulsar emission and polarization between the leading and trailing sides. For example the coherent emitters may be brighter on one side of the profile than on the other side depending upon the factors influencing the coherence process. And this can act in the same direction or in opposite direction to the processes that we have considered in this paper.

Note that we have performed modeling of coherent pulsar radio emission in terms of single particle curvature radiation. Although to the first order it is not a bad assumption, in reality the efficiency of coherence process may depend on many factors and hence there may be an induced asymmetries in the pulsar emission and polarization between the leading and trailing sides. For example the coherent emitters may be brighter on one side of the profile than on the other side depending upon the factors influencing the coherence process. And this can act in the same direction or in opposite direction to the processes that we have considered in this paper.

\section{CONCLUSION}
We believe our model is much more realistic for pulsar emission and
polarization than those proposed before, as for the first time it
simultaneously takes into account of the perturbation caused jointly by
the rotation and the PC--current. In addition it considers the detailed study
on nonuniform source distribution (modulation) and the influence of
viewing geometry on pulsar emission. Hence we believe one can explain
most of the observed emission and polarization properties of pulsars
within the frame work of curvature radiation. On the basis of our
simulation of pulse profiles we draw the following conclusions:
\begin{enumerate}
\item The effect of rotation and PC--current perturbation in the
  presence of nonuniform source distribution (modulation) along with
  viewing geometry might be responsible for the most of the observed
  diverse behavior of polarization properties of pulsar radio
  emission.
\item Because of the perturbations there arises an asymmetry in the
  phase shift of intensity components and PPA inflection point for
  $\pm\sigma.$ Further we notice that the phase delay of the PPA
  inflection point with respect to that of the central component
  (core) becomes larger for negative $\sigma$ than that for the
  positive $\sigma$.
\item The leading side components can become either stronger or weaker
  than the corresponding trailing side components of a given
  cone. This is due to the induced asymmetry in the curvature of
  source trajectory and the sight line encountered asymmetry in the
  modulation strength between the two sides.
\item Both the ``antisymmetric'' and ``symmetric''$-$types circular
  polarization are possible within the frame work of curvature
  radiation when the perturbation and modulation are operative.
\item In the presence of perturbation the sign reversal of the
  ``antisymmetric''$-$type circular polarization as well as the sign
  of the ``symmetric''$-$type circular polarization do not show the
  correlation with the sign of the PPA swing in the case of central
  core dominated pulsars.
\item The `kinky' type distortions in the PPA swing could be due to
  the incoherent addition of modulated emissions in presence of strong
  perturbations.
\end{enumerate}
\clearpage 
\end{document}